\documentclass[reprint,amsmath,amssymb,aps,prd,longbibliography]{revtex4-1}

\usepackage{amsmath}
\usepackage{graphicx}
\usepackage{bm}
\usepackage{bbm}
\usepackage{braket}
\usepackage{tensor}
\usepackage{tikz}
\usepackage{pgfplots}
\usetikzlibrary{matrix,hobby}
\pgfplotsset{compat=1.13}

\begin{document}

\title{Tidal and nonequilibrium Casimir effects in free fall}

\author{Justin H.\ Wilson$^{1}$}
 \email{justin@jhwilson.com}
 \affiliation{
Department of Physics and Astronomy, Center for Materials Theory, Rutgers University, Piscataway, New Jersey 08854 USA
}
\affiliation{%
Institute of Quantum Information and Matter and Department of Physics,
Caltech, Pasadena, California 91125
}%
\author{Francesco Sorge$^{2}$}
 \email{sorge@na.infn.it}
\affiliation{I.N.F.N. - Complesso Universitario Monte S. Angelo, via Cintia, Ed. 6,
80126 Napoli, Italy
}

\author{Stephen A.\ Fulling$^{3}$}
\email{fulling@math.tamu.edu}
\affiliation{$^3$ Department of Mathematics and Institute for Quantum Science and Engineering, Texas A\&M University, College Station, Texas, 77843-3368}

\date{\today}

\begin{abstract}
  In this work, we consider a Casimir apparatus that is put into free fall (e.g., falling into a black hole).
  Working in 1+1D, we find that two main effects occur: First, the Casimir energy density experiences a tidal effect where negative energy
  is pushed toward the plates and the resulting force experienced by the plates is increased.
  Second, the process of falling is inherently nonequilibrium and we treat it as such, demonstrating that the Casimir energy density moves back and forth between the plates after being ``dropped,'' with the force modulating in synchrony.
  In this way, the Casimir energy behaves as a classical liquid might, putting (negative) pressure on the walls as it moves about in its container. 
  In particular, we consider this in the context of a black hole and the multiple vacua that can be achieved outside of the apparatus.
\end{abstract}


\maketitle

\section{Introduction}

The Casimir effect in flat space causes two distinct objects (such as plates, as Casimir originally considered \cite{casimirAttractionTwoPerfectly1948a}) to attract with a pressure that is diminished as the objects recede (see, e.g., \cite{bordagNewDevelopmentsCasimir2001}).
There are two competing ways of conceptualizing the force, one based on the van der Waals picture of fluctuating dipole moments interacting through the photon field, and the other on the QED vacuum energy changes between the plates.
These conceptualizations are not in conflict; in principle they are two different ways of doing the same energy bookkeeping, though in practice the experimental regimes where each is useful are complementary \cite{rodriguezCasimirEffectMicrostructured2011}.
In this work we concentrate on the vacuum-energy picture, in a highly idealized model.
As we will show, the energy density between the objects behaves like a fluid: it is subject to both tidal forces and nonequilibrium effects (including the so-called dynamical Casimir effect \cite{dalvitFluctuationsDissipationDynamical2011a,nationColloquiumStimulatingUncertainty2012,wilsonObservationDynamicalCasimir2011}).

The subject of quantum field theory in a one-dimensional moving cavity (hence, two accelerating, perfectly reflecting boundaries) was initiated by Moore \cite{mooreQuantumTheoryElectromagnetic1970}.
Independently, DeWitt \cite{DeWitt1975} studied the effect of a single accelerating boundary, which provides the foundation of the  dynamical part of the effects predicted by Moore.
The theory was further developed in a series of papers \cite{fullingRadiationMovingMirror1976,daviesEnergymomentumTensorEvaporating1976,daviesQuantumVacuumEnergy1977,daviesRadiationMovingMirrors1977}, the last three of which moved into the context of curved space-time.
The most general situation thus combines three ingredients:  space-time curvature (possibly time-dependent), moving boundaries (causing the particle creation  now commonly
known as dynamical Casimir effect), and a cavity of finite size (creating the vacuum energy that generalizes the true, static Casimir effect).
We use the general theory in Davies--Fulling \cite{daviesQuantumVacuumEnergy1977} as a starting point for our analysis.  There, however, little attention was paid to the combination of curvature with finite size, which is the primary concern of the present paper.

Since the 1970s, the study of the dynamical Casimir effect in a cavity has largely developed independently of its curved-space origins \cite{dodonovNonstationaryCasimirEffect2002,dodonovQuantumHarmonicOscillator2005}.
The dynamical Casimir effect has attracted interest with analogies in superconducting circuits \cite{nationColloquiumStimulatingUncertainty2012} that have seen experimental verification \cite{wilsonObservationDynamicalCasimir2011}.
Developments that include effects of the discrete cavity spectrum include inverse solutions to Moore's original equation \cite{castagninoRadiationMovingMirrors1984}, non-relativistic perturbative solutions \cite{dodonovNonstationaryCasimirEffect1989}, solutions with constant relative velocity \cite{bordagCalculationCasimirEffect1984,bordagCasimirEffectUniformly1986}, and vibrating boundaries \cite{fordVacuumEnergyDensity1998,dalvitRenormalizationgroupApproachDynamical1998,dalvitCreationPhotonsOscillating1999} amongst others.
We now add to this list an apparatus in free fall.

In curved space, the photon propagator changes between the plates and so one expects the Casimir effect to be affected.
The Davies--Fulling papers considered the exactly solvable situation of a massless scalar field in two-dimensional space-time.
The qualitative picture is similar for (e.g.)\ four-dimensional electrodynamics, but explicit calculations are much harder.

\begin{figure}
    \centering
    \includegraphics[width=\columnwidth]{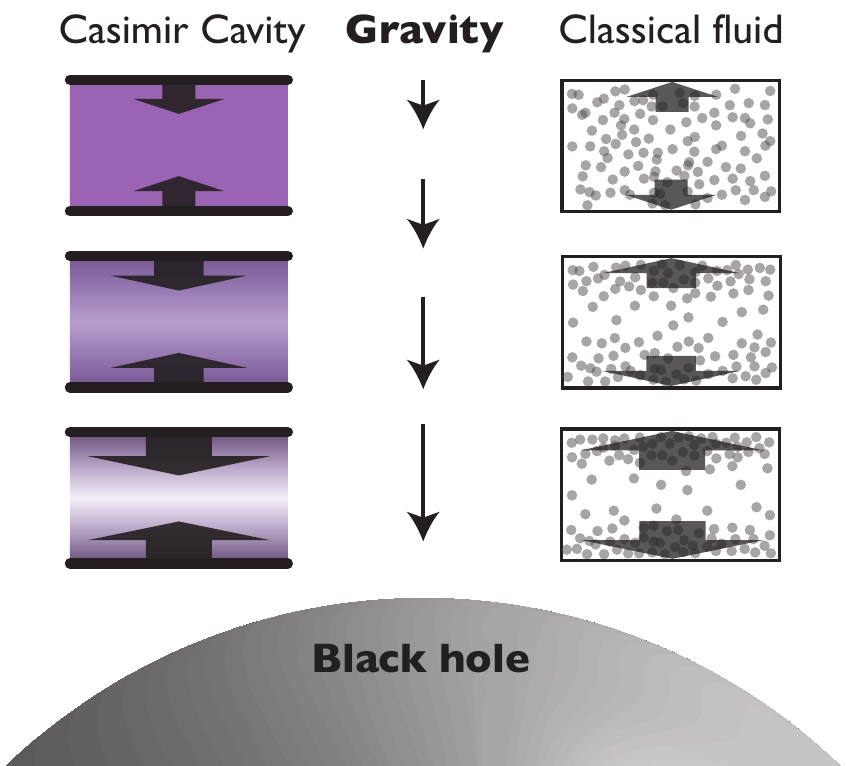}
    \caption{Just as particles in a falling box experience a tidal force which forces them to the sides (right), the negative Casimir energy experiences a similar tidal force (left). The result, much like the box, is an increase in (negative) Casimir pressure on the two plates. Additionally, a falling Casimir apparatus is inherently dynamical, and excitations will be created (not pictured) both outside and inside the box.}
    \label{fig:my_label}
\end{figure}

Consider a Casimir apparatus (precisely, a 1+1D scalar field theory in a cavity) where two plates are kept a fixed distance, $L$, from one another and their center of mass falls along a geodesic.
To understand how the force can be modified by curvature, it is useful to consider the analogous system of a box full of fluid as in Fig.~\ref{fig:my_label}.
In thermodynamic equilibrium, the gas exerts a pressure on the walls of the box, equal on all sides.
Now, consider that we take this box and drop it; two things should happen.
First, the particles will be put out of equilbrium and will ``slosh'' around the box causing a different pressure at different times.
Second, tidal effects will push some of the gas to either side of the box, effectively causing an added pressure on those walls.
Now return to the  Casimir apparatus.
At rest in flat space it exhibits a normal Casimir effect and the (regularized) energy between the plates is flat (constant) between them.
 We shall show that when we drop this apparatus two things occur.
First, the system is thrust out of equilibrium, as one can see explicitly from the energy density, which begins to change in time (it ``sloshes'' back and forth with period $2L/c$).
Second, the energy density exhibits a \emph{tidal} effect whereby negative energy density moves toward the plate.
While the total Casimir energy is less than in free space in magnitude, the pressure on the plates increases.
So far, it is in direct analogy with particles;
however, this analogy does break down when we consider that particle number is not conserved and, quite generally, moving plates will create excitations that will contribute to the energy density.
Nonetheless, those dynamical terms are easily identified and characterized, as we will see.

In Sec.~\ref{sec:prelim} we review the basic theory and what is already known about this problem.
Then, in Sec.~\ref{sec:GeneralTheory} we develop a general theory to handle two moving plates, and also a
perturbation theory for detailed study of a free-fall Casimir apparatus with fixed proper distance between the plates.
In Sec.~\ref{sec:EnergyMomentumTensor} we calculate all relevant terms in the energy-momentum tensor inside and outside of the Casimir apparatus.
And in Sec.~\ref{sec:CasimirEffect} we fully explore the force on the plates and the energy density between them --- observing explicitly the tidal and nonequilibrium Casimir effects.
Lastly, in Sec.~\ref{sec:blackhole} we apply all of this to the case of a Casimir apparatus falling into a black hole and find that the Casimir attraction between the plates increases from both the tidal effects and the dynamical effects.

Throughout this work, we use the standard conventions $\hbar = 1 = c$, and for consistency with the 1970s literature the metric signature is ($+$ $-$) (i.e., the minus sign is associated with the spatial dimension).

\section{Preliminaries}
\label{sec:prelim}

We are considering a 1+1D scalar field theory defined by the action
\begin{equation}
  S = \int d^2 x \sqrt{-g}\, g^{\mu \nu} \partial_\mu\phi \,\partial_\nu \phi,
\end{equation}
where $\phi$ is the field and $g_{\mu\nu}$ is the 
metric tensor.

The line element (and hence metric) in $1+1$D can always be written --- nonuniquely --- in the conformally flat form
\begin{equation}
    ds^2 = C(u,v)\, du\, dv, \label{eq:metricOnePlusOne}
\end{equation}
where $(u,v)$ are null coordinates ($u = t-x$ and $v = t+x$ where $t$ and $x$ are respectively timelike and spacelike coordinates), and $C(u,v)$ is the conformal factor.
The wave equation in these coordinates is simply $\partial_u \partial_v \phi = 0$.
These coordinates also imply a natural set of Cauchy surfaces defined by the timelike vector field $\partial_t \equiv \frac12 (\partial_u + \partial_v)$.
Quantizing the field with these surfaces gives us positive energy modes, satisfying $i\partial_t \phi = \omega\phi$ with $\omega>0$.

The only nonzero Christoffel symbols for this metric are
\begin{equation}
  \begin{split}
  \Gamma\indices{^u_{uu}} & = \partial_u \log C, \\ \Gamma\indices{^v_{vv}} & = \partial_v \log C
\end{split}
\end{equation}
and the Ricci curvature scalar (which, in 1+1D, completely determines the geometry locally) is
\begin{equation}
  R = -4 \frac{\partial_v\partial_u \log C}C = -\Box \log C. \label{eq:RicciScalar}
\end{equation}

A key observation present in the very early works \cite{mooreQuantumTheoryElectromagnetic1970,DeWitt1975,fullingRadiationMovingMirror1976,daviesQuantumVacuumEnergy1977,birrellQuantumFieldsCurved1984} on the subject is that the subset of conformal transformations $u\rightarrow \bar u = f(u)$ and $v \rightarrow \bar v = g(v)$ leave the metric conformally flat but modify the conformal factor:
\begin{equation}
  C(u,v) \rightarrow \bar C(\bar u, \bar v) = \frac{C(u,v)}{f'(u) g'(v)}\,; \label{eq:transformC}
\end{equation}
then the physics, boundary conditions, and causal structure dictate how to choose $f$ and $g$.

In order to define a ``vacuum'' state, we need a timelike vector field; it defines a set of Cauchy surfaces upon which we can write a Hamiltonian operator $H$ and hence arrive at a preferred vacuum-like state $\ket 0$.
This vector field is conveniently encoded in the coordinates we use by $\partial_t = \frac12(\partial_u + \partial_v)$, so a ``conformal coordinate transformation'' is equivalent to picking a new vector field with which to define $H$; this gives the construction a more intrinsic geometrical flavor, as $-C$ is the norm of the vector field and $R$ can be rewritten as in the right member of (\ref{eq:RicciScalar}) \cite{BCG2012}.
In general, different vector fields give different states $\ket 0$.

That difference  is highlighted by the expectation value (in $\ket0$) of the energy-momentum tensor \cite{daviesQuantumVacuumEnergy1977},
the formula for which is
\begin{equation}
  \braket{T_{\mu \nu}} = \theta_{\mu \nu} - \frac1{48\pi} R g_{\mu \nu}, \label{eq:EnergyMomentumTensor}
\end{equation}
where, in the case of two Dirichlet plates [$\phi(0)=0=\phi(L)$] separated by a coordinate distance $L$,
\begin{align}
  \theta_{uu} & = \frac1{24\pi} F_u(C) - \frac{\pi}{48 L^2}\,, \label{eq:thetauu} \\
  \theta_{vv} & = \frac1{24\pi} F_v(C) - \frac{\pi}{48 L^2}\,, \label{eq:thetavv} \\
  \theta_{uv} & = \theta_{vu}=0, \label{eq:thetauv}
\end{align}
with
\begin{equation}
  F_x(f) = \frac{f''(x)}{f(x)} - \frac32\left(\frac{f'(x)}{f(x)} \right)^2.
\end{equation}

 The quantity $\theta_{\mu\nu}$ naturally breaks up into two terms,   obtained from Eqs.~\eqref{eq:thetauu} and \eqref{eq:thetavv} as
\begin{align}
  \theta^{\mathrm{dyn}}_{u u} & = \frac1{24\pi } F_{u} ( C), &
   \theta^{\mathrm{dyn}}_{v v} & = \frac1{24\pi } F_{v} ( C), \label{zydyn} \\
   \theta^{\mathrm{stat}}_{ u u} & = -\frac{\pi}{48 L^2}\,, & \theta^{\mathrm{stat}}_{ v v} &= -\frac{\pi}{48 L^2}\,.
\label{zystat}\end{align}
The dynamical term \eqref{zydyn} originates as radiation (particle creation)  from each individual plate (the DeWitt effect); later this radiation suffers
reflections from both plates.  This term, which depends essentially on the time dependence of the geometry, corresponds most directly to what is called ``dynamical Casimir effect'' in the recent literature,
but is more legitimately called the Moore effect.
The other term, \eqref{zystat}, we call \emph{quasistatic};
it is the direct descendent
of the usual Casimir effect of static plates in flat space (caused by
discreteness of the mode spectrum), and
its existence  requires the presence of more than one plate.
Section~\ref{sec:EnergyMomentumTensor} demonstrates that both the tidal and the sloshing effects mentioned previously are exhibited by the quasistatic term, in the presence of curvature and time dependence, respectively.

Note that $F_x(f')$ is exactly the Schwarzian derivative \cite{taborChaosIntegrabilityNonlinear1989} of $f$;
this highlights the connection between 2D Lorentzian geometry and complex analysis.
When the argument of $F_{u}$ or $F_v$ is the conformal factor, the result can be written solely in terms of the Christoffel symbols:
\begin{align}
  F_u(C) & = \partial_u \Gamma^u_{uu} - \tfrac12 \Gamma^u_{uu}, \label{eq:FuChristofell}\\
  F_v(C) & = \partial_v \Gamma^v_{vv} - \tfrac12 \Gamma^v_{vv}. \label{eq:FvChristofell}
\end{align}

To highlight how the appearance of Christoffel symbols indicates dependence on the vector field $\partial_t$, note that $F_u(C)$ has nontrivial transformation properties. In particular, following Eq.~\eqref{eq:transformC}
\begin{equation}
  F_{\bar u}(\bar C) = \frac1{f'(u)^2}\left[F_u(C) - F_u(f')\right].
\end{equation}
It looks as though $\braket{T_{\mu\nu}}$ is not transforming like a tensor, but in fact there are two different, tensorial, stress tensors involved,
because $\ket{\bar 0}$, the vacuum defined with respect to $\partial_{\bar t}$, is not the same as $\ket 0$, defined by $\partial_t\,$.
For Eqs.~\eqref{eq:thetauu}--\eqref{eq:thetauv}, the Cauchy surfaces are defined by the vector field $\partial_t\,$.

We can
 make a series of coordinate transformations of the $ (f(u),g(v))$ type to put both of our plates at fixed coordinate positions.
In such coordinates the field equation  is easily solved by d'Alembert's construction.
However, the set of coordinate transformations that can do this is not unique (and hence we have inequivalent vector fields with different vacuum states), so a central technical result of this paper is when and how to determine a unique coordinate transformation and hence a unique initial vacuum.

\section{General theory for two moving plates}\label{sec:GeneralTheory}
\subsection{The general construction}\label{sec:General}
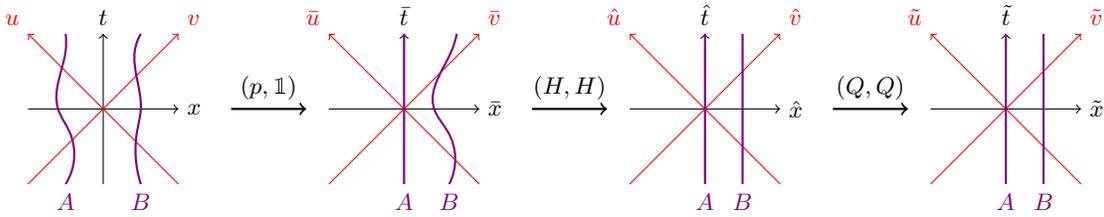
\begin{figure*}
  \begin{center}
  \begin{tikzpicture}

   \foreach \x/\xname/\tname/\uname/\vname in {0/$x$/$t$/$u$/$v$,%
                                               4/$\bar x$/$\bar t$/$\bar u$/$\bar v$,%
                                               8/$\hat x$/$\hat t$/$\hat u$/$\hat v$,%
                                               12/$\tilde x$/$\tilde t$/$\tilde u$/$\tilde v$}{%
   \draw[->] ({-1+\x},0) -- ({1+\x},0) node(xline)[right] {\xname};
   \draw[->] ({0+\x},-1) -- ({0+\x},1) node(yline)[above] {\tname};
   \draw[color=red,->] ({-1+\x},-1) -- ({1+\x},1) node [anchor=south west] {\vname};
   \draw[color=red,->] ({1+\x},-1) -- ({-1+\x},1) node [anchor=south east] {\uname};}

   \foreach \x/\labarr in {0/${(p,\mathbbm 1)}$,%
                           4/${(H,H)}$,%
                           8/${(Q,Q)}$}%
                           {%
   \draw[->,thick] ({1.7+\x},0) -- node[above] {\labarr} ({2.7+\x},0);}

   \draw[color=violet,thick] (-0.5,-1) node[below] {$A$} to [curve through ={(-0.45,-0.3) .. (-0.6,0) .. (-0.5,0.77)}] (-0.5,1);
   \draw[color=violet,thick] (0.5,-1) node[below] {$B$} to [curve through ={(0.45,-0.3) .. (0.5,0) ..(0.42,0.77)}] (0.5,1);

   \draw[color=violet,thick] (4,-1) node[below] {$A$} -- (4,1);
   \draw[color=violet,thick] (4.6,-1) node[below] {$B$} to [curve through ={(4.58,-0.3) .. (4.4,0) ..(4.63,0.77)}] (4.7,1);

   \draw[color=violet,thick] (8,-1) node[below] {$A$} -- (8,1);
   \draw[color=violet,thick] (8.5,-1) node[below] {$B$} -- (8.5,1);

   \draw[color=violet,thick] (12,-1) node[below] {$A$} -- (12,1);
   \draw[color=violet,thick] (12.5,-1) node[below] {$B$} -- (12.5,1);

  \end{tikzpicture}
  \end{center}
  \caption{The series of coordinate transformations made to define coordinates where the plates' coordinates are at constant coordinate positions. The map $(p,\mathbbm 1)$ represents the mapping $(u,v) \mapsto (p(u),v)$ (and similarly for the other mappings). We first ``straighten'' out plate $A$ with $(p,\mathbbm 1)$, then plate $B$ with $(H,H)$. After that, there is an infinite-dimensional space of solutions that keep the plates at constant coordinate position but are nontrivial transforms $(Q,Q)$. In an initially static model a unique $Q$ can be determined by causality  as explained in Fig.~\ref{fig:InitialCond_plates} and associated text. \label{fig:coordinate-transformations}}
\end{figure*}

We begin by considering two plates labeled by $A$ and $B$ on arbitrary trajectories in spacetime.
By convention, $B$ is to the right of $A$ (larger $x$ at any given~$t$).
Written in null coordinates the trajectories $P_A$ and $P_B$ are
\begin{equation}
  P_A(\tau) = (U_A(\tau), V_A(\tau)), \quad P_B(\tau)  = (U_B(\tau), V_B(\tau)),
\end{equation}
where $\tau$ is for the moment just a parameter, not necessarily proper time.
The trajectories represent boundary conditions on $\phi$ which we take to be perfectly reflecting:
\begin{equation}
  \phi(P_A) = 0 = \phi(P_B).
\end{equation}
This problem becomes simple to solve (i.e., Eqs.~\eqref{eq:thetauu}--\eqref{eq:thetauv} directly apply) if we can transform the trajectories to be on constant spatial coordinates by appropriate conformal transformations (of the type $(u,v) \mapsto (f(u),g(v))$).
We will need two coordinate transformations to put the plates on constant coordinates; then we will investigate degenerate mappings that keep the plates at fixed coordinates.
The series of transformations will be indicated by
\begin{equation}
  (u,v)\mapsto (\bar u, \bar v) \mapsto (\hat u, \hat v) \mapsto (\tilde u, \tilde v)
\end{equation}
and are pictured in Fig.~\ref{fig:coordinate-transformations}.

The first coordinate transformation  puts $P_A$ on a constant coordinate:
\begin{equation}
  (u,v) \longmapsto (\bar u, \bar v) = (p(u), v), \quad p = V_A \circ U_A^{-1}.
\end{equation}
This is represented by the mapping $(p,\mathbbm 1)$ in Fig.~\ref{fig:coordinate-transformations}.
As we can see, $(U_A, V_A)\mapsto (V_A, V_A)$, indicating that the plate is at $\bar x \equiv \frac12(\bar u - \bar v) = 0 $.
Causality is already playing an important role in how this coordinate transformation is chosen:
If the plate begins moving at $U_A=0=V_A$, for instance, then only space-time points with $u>0$ have light rays
which carry information that the plate has started to move.
Therefore, any change in the $v$ coordinate here would give an acausal vacuum.

Now, we become more constrained by the process of putting the second plate on a fixed coordinate.
First, to keep the plate $A$ in the same spot, we need
\begin{equation}
  (\bar u, \bar v) \longmapsto (\hat u, \hat v) = (H(\bar u), H(\bar v)),
\end{equation}
but we also would like to put plate $B$ at a fixed \emph{coordinate} distance $L$ (which will often be set equal to proper distance without loss of generality), which leads us to the constraint
\begin{equation}
  H\circ V_B(\tau) - H\circ p \circ U_B(\tau) = 2L\,;
\label{eq:Hagain}\end{equation}
these conditions give the coordinate transformation $(H,H)$ in Fig.~\ref{fig:coordinate-transformations}.

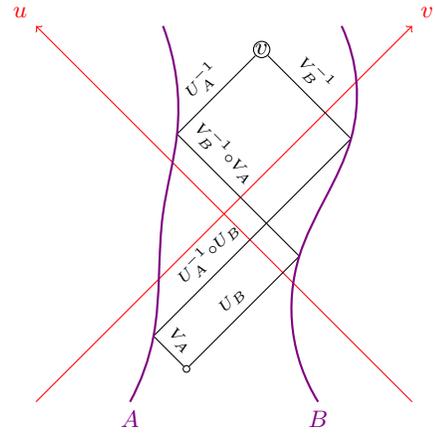
\begin{figure}
 \begin{tikzpicture}[scale=1.25]
   \draw[color=red,->] (-2,-2) -- (2,2) node [anchor=south west] {$v$};
   \draw[color=red,->] (2,-2) -- (-2,2) node [anchor=south east] {$u$};

   \def\initx{0.4}
   \def\inity{1.75}
   \def\rightOne{0.95}
   \def\leftOne{0.9}
   \def\rightTwo{2.1}
   \def\leftTwo{1.3}

   \node[draw,circle,inner sep=0pt, minimum size=2.5pt] (A) at (\initx,\inity) {$v$};
   \node[inner sep=0pt,minimum size=0pt] (BR) at ({\initx+\rightOne},{\inity-\rightOne}){};
   \node[inner sep=0pt,minimum size=0pt] (BL) at ({\initx-\leftOne},{\inity-\leftOne}){};
   \node[inner sep=0pt,minimum size=0pt] (CR) at ({\initx+\rightOne-\rightTwo},{\inity-\rightOne-\rightTwo}){};
   \node[inner sep=0pt,minimum size=0pt] (CL) at ({\initx-\leftOne+\leftTwo},{\inity-\leftOne-\leftTwo}){};

   \node[draw, circle, inner sep=0pt, minimum size=2.5pt] (D) at ({\initx-\rightTwo+\leftTwo},{\inity-\leftTwo-\rightTwo}){};

   \draw (A) -- (BR) node [pos=0.4,above,sloped]{$\scriptstyle V_B^{-1}$} -- (CR) %
          node [pos=0.65,above,sloped]{$\scriptstyle U_A^{-1}\circ U_B$} -- (D)%
          node [pos=0.5,above,sloped]{$\scriptstyle V_A$};
   \draw (A) -- (BL) node [pos=0.5,above,sloped]{$\scriptstyle U_A^{-1}$} -- (CL) %
          node [pos=0.27,above,sloped]{$\scriptstyle V_B^{-1} \circ V_A$} -- (D)%
          node [pos=0.5,above,sloped]{$\scriptstyle U_B$};

   \draw[color=violet,thick] (-1,-2) node[below] {$A$} to [curve through ={(CR) .. (-0.65,0) .. (BL)}] (-0.65,2);
   \draw[color=violet,thick] (1,-2) node[below] {$B$} to [curve through ={(CL) .. (BR)}] (1.25,2);

 \end{tikzpicture}
 \caption{This shows how the equations \eqref{eq:H_solve} and \eqref{eq:Hcircp_solve} associate a coordinate at one point in space to another point by reflections off both mirrors. The operations should be read top to bottom. This describes the coordinate transformation $H$ that puts the plates on constant coordinates after we transform $(u,v)\mapsto (\hat u,\hat v)=(H\circ p(u), H(v))$.\label{fig:nullcoordinate_COV}}
\end{figure}

We can rewrite Eq.~\eqref{eq:Hagain} as
\begin{equation}
  H(v)= H\circ V_A \circ U_A^{-1}\circ U_B \circ V_B^{-1}(v) + 2L, \label{eq:H_solve}
\end{equation}
and also as
\begin{equation}
  (H\circ p)(u) = (H\circ p)\circ U_B \circ V_B^{-1}\circ V_A \circ U_A^{-1}(u) + 2L. \label{eq:Hcircp_solve}
\end{equation}
To understand Eqs.~\eqref{eq:H_solve} and \eqref{eq:Hcircp_solve} and what motivates them, consider Fig.~\ref{fig:nullcoordinate_COV}.
Essentially, $H(v)$ is equal to the coordinate one gets by tracing a light ray back to plate $B$, reflecting to plate $A$ and registering the corresponding $v$ coordinate there (and similarly for $H\circ p$).
In this light-ray tracing scheme, every point in space is related to a point in its past after two reflections, and that defines a coordinate transformation that puts the plates at constant spatial coordinates.

However, the solutions to Eq.~\eqref{eq:H_solve} are not unique.
Consider a transformation
\begin{equation}
  (\hat u, \hat v) \longmapsto (\tilde u, \tilde v) = (Q(\hat u), Q(\hat v))
\end{equation}
such that plates $A$ and $B$ both remain at constant coordinate position.
In that case, the only condition on $Q$ is
\begin{equation}
  Q(\hat v) = Q(\hat v - 2L) + 2L. \label{eq:Q_solve}
\end{equation}
This equation has multiple solutions, as can be easily seen if one lets $Q(\hat v) = \hat v + \Omega(\hat v)$ with $\Omega(\hat v) = \Omega(\hat v - 2L)$.
Thus $\Omega$ is any periodic function with period $2L$ --- therefore, we have a whole continuum of solutions.
The final transformation from $(\bar u, \bar v) \mapsto (\tilde u, \tilde v)$ is then
\begin{equation}
  \tilde H = Q \circ H.
\end{equation}

While this function $\Omega$ is an interesting mathematical oddity, it has real physical consequences.
First, we can guess that the periodicity of $2L$ represents a sloshing of the field between the plates, and is therefore an indication that we are probing some out-of-equilibrium phenomenon.
To be precise, one can solve for the normal modes between the plates and get two sets of functions (neglecting normalization),
\begin{equation}
  \begin{split}
  \hat \psi_n(\hat t, \hat x) & = e^{-i\omega_n \hat t} \sin(\omega_n \hat x), \\
  \tilde \psi_n(\tilde t, \tilde x) & = e^{-i\omega_n \tilde t} \sin(\omega_n \tilde x) ,
\end{split}
\end{equation}
where $\omega_n = n \pi/L$ ($n>0$) and $(\hat t,\hat x)$ are defined by $\hat u = \hat t - \hat x$ and $\hat v = \hat t + \hat x$ (and similarly for the tilde-coordinates).
These functions have positive energy with respect to their vector fields: $i\partial_{\hat t} \hat \psi_n = \omega_n \hat \psi_n$ and $i\partial_{\tilde t} \tilde \psi_n = \omega_n \tilde \psi_n$ with $\omega_n>0$.

Now, if we make the substitution that $\tilde u = \hat u + \Omega(\hat u)$ and $\tilde v = \hat v + \Omega(\hat v)$, we first notice
\begin{equation}
  \tilde \psi_n = \frac1{2i}\left[ e^{-i\omega_n \hat u - i \omega_n\Omega(\hat u)} - e^{-i \omega_n \hat v - i \omega_n \Omega(\hat v)} \right],
\end{equation}
and taking advantange of the periodicity in $\Omega$, we expand
\begin{equation}
e^{-i\omega_n \Omega(\hat w)} = \left[\alpha_0 + \sum_{m=1}^\infty(\alpha_{nm} e^{-i \omega_m \hat w} + \beta_{nm} e^{i \omega_m \hat w})\right] e^{i\omega_n \hat w}
\end{equation}
to obtain
\begin{equation}
  \tilde \psi_n = \sum_{m=1}^\infty \left[ \alpha_{nm} \hat \psi_m(\hat t, \hat x) + \beta_{nm} \hat \psi^*_m(\hat t, \hat x)\right].
\end{equation}
By standard techniques \cite{birrellQuantumFieldsCurved1984} this can be used to relate annihilation operators ($\hat a_m$ for the hatted-coordinates and $\tilde a_n$ for the tilde-coordinates) by
\begin{equation}
 \hat a_m = \sum_{n}[ \alpha_{mn} \tilde a_n + \beta_{mn}^*\tilde a^\dagger_n],
\end{equation}
so the vacuum $\ket{\tilde 0}$ defined by $\tilde a_n \ket{\tilde 0}=0$ is not annihilated by $\hat a_m\,$, and the number of excitations in the hatted coordinates is
\begin{equation}
 \braket{\tilde 0 | \hat a_m^\dagger \hat a_m | \tilde 0} = \sum_n | \beta_{mn} |^2.
\end{equation}
For even simple periodic functions $\Omega$ this quantity is nonvanishing.

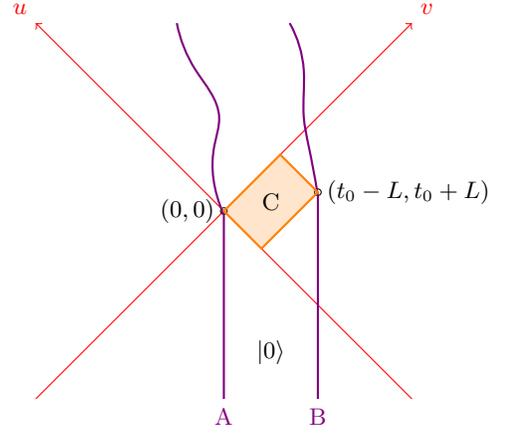
\begin{figure}
 \begin{tikzpicture}[scale=1.25]
   \draw[color=red,->] (-2,-2) -- (2,2) node [anchor=south west] {$v$};
   \draw[color=red,->] (2,-2) -- (-2,2) node [anchor=south east] {$u$};

   \node[draw,circle,inner sep=0pt, minimum size=2.5pt] (A) at (0,0) {};
   \node[draw,circle,inner sep=0pt, minimum size=2.5pt] (B) at (1,0.2) {};

   \draw[color=violet,thick] (0,-2) node[below] {A} -- (A) node[left,color=black] {$(0,0)$} to [curve through ={(-0.1,0.7) .. (-0.05,1) .. (-0.3,1.5)}] (-0.5,2);
   \draw[color=violet,thick] (1,-2) node[below] {B} -- (B) node[right,color=black]{$(t_0-L,t_0+L)$} to [curve through ={(0.9,0.7) .. (0.85,1) .. (0.85,1.5)}] (0.7,2);

   \filldraw [thick,draw=orange,fill=orange,fill opacity=0.2] (A.center) -- (0.6,0.6) -- (B.center) -- (0.4,-0.4) -- cycle;



   \node at (0.5,-1.5) {$\ket{0}$};
   \node at (0.5,0.1) {C};

 \end{tikzpicture}
 \caption{In order to get physically sensible results, in the past of region C we assume our space-time is static, and therefore we can define a coordinate system in which the plates are initially at rest with a vacuum $\ket{0}$ defined with respect to the vector field $\partial_t\,$.
 (Imagine dropping plates from above a black hole as we will consider in Sec.~\ref{sec:blackhole}.)  Then at $(0,0)$ [i.e., $t=0$] plate $A$ begins to move and at $(t_0-L, t_0+L)$ [i.e., $t=t_0$] plate $B$ begins to move. We can solve Eq.~\eqref{eq:H_solve} using these initial conditions.
 On the other hand, if we solve Eq.~\eqref{eq:H_solve} in the future of region C without specifying the initial conditions, we can use the function $Q$ in Eq.~\eqref{eq:Q_solve} to implement those initial conditions. The result is Eqs.~\eqref{eq:OmegaNeg} and \eqref{eq:OmegaPos}. \label{fig:InitialCond_plates}}
\end{figure}

Now, we can identify this function $\Omega$ with the initial conditions of the plates.
To prove this, consider the physical situation given in Fig.~\ref{fig:InitialCond_plates}.
At $t=0$ plate $A$ begins to move and at $t=t_0$ with $|t_0|<L$ plate $B$ begins to move.
In the region in the past of C as pictured in Fig.~\ref{fig:InitialCond_plates}, the vector field that defines the vacuum state is $\partial_t$.
This defines our initial conditions.

If we just solve Eq.~\eqref{eq:H_solve} we may not satisfy the initial conditions we desire.
To find a $Q$ and hence a $\tilde H$ such that $\ket{0}$ is the initial vacuum state, we can look at region C in Fig.~\ref{fig:InitialCond_plates}, where by causality no information about the moving plates exists. Thus, we impose
\begin{align}
   \tilde H \circ p(u) &= u, & t_0-L < u < 0, \label{eq:u-initialconds} \\
   \tilde H(v) &= v, & 0<v < t_0+L. \label{eq:v-initialconds}
\end{align}
We can now determine the form of $\Omega$ over all $\hat{v}$ by induction.
First note that under the arbitrary $H$ that solves Eq.~\eqref{eq:H_solve}, $(t_0-L,t_0+L)\stackrel{H}{\longmapsto}(t_1 - L, t_1 + L)$ for some $t_1$ and $(0,0)\stackrel{H}{\longmapsto}(0,0) $.
Therefore, we can take Eq.~\eqref{eq:u-initialconds} along with $\tilde H = Q \circ H$ to determine
\begin{align}
  H\circ p(u) &+ \Omega[H\circ p(u)]  = u, & t_0 - L < u < 0, \\
  \Omega(w) & = (H\circ p)^{-1}(w) - w, & t_1 - L < w < 0. \label{eq:OmegaNeg}
\end{align}
Then, using Eq.~\eqref{eq:v-initialconds} we have
\begin{align}
  H(v) & + \Omega[H(v)]  = v, & 0 < v < t_0 + L, \\
  \Omega(w) & = H^{-1}(w) - w, & 0 < w < t_1 + L. \label{eq:OmegaPos}
\end{align}
Equations \eqref{eq:OmegaNeg} and \eqref{eq:OmegaPos} specify $\Omega$ over a range of $2L$, so they uniquely specify our periodic function.
The function is also continuous.
To show this,  note first that $\Omega(0^-) = 0 = \Omega(0^+)$, and second that
\begin{equation}
  \Omega(t_1 + L)  = t_0 - t_1  = \Omega(t_1 - L).
\end{equation}
Finally, we introduce the notation $\{X\}_{t_1 - L}^{t_1 + L}$ defined by
\begin{align}
  X = 2 n L + \{X\}_{t_1 - L}^{t_1 + L}, & \quad n\in \mathbb Z, \\ &  t_1 - L < \{X\}_{t_1 - L}^{t_1 + L} < t_1 + L,\nonumber
\end{align}
so that we can write the full solution as
\begin{multline}
  \tilde H(v) = H(v) - \{H(v)\}_{t_1 - L}^{t_1 + L} \\ + \begin{cases}
  (H\circ p)^{-1}[\{H(v)\}_{t_1 - L}^{t_1 + L}], & \{H(v)\}_{t_1 - L}^{t_1 + L} < 0, \\
  H^{-1}[\{H(v)\}_{t_1 - L}^{t_1 + L}], & \{H(v)\}_{t_1 - L}^{t_1 + L} > 0.
  \end{cases} \label{eq:Htilde_soln}
\end{multline}
With this general analysis, we can solve Eq.~\eqref{eq:H_solve} with a method that produces an arbitrary solution $H$, then using Eq.~\eqref{eq:Htilde_soln} we can find $\tilde H$.

Given this understanding of what the ``correct'' vacuum state is,  it is important to understand that the state in
a space-time region such as region C displayed in Fig.~\ref{fig:InitialCond_plates} is not completely determined by the plate's motion displayed there.  The plates might not be static everywhere in the past of that region;
there might be an earlier period of wiggling, off the bottom of the figure.  In that case the functions $\Omega$, $Q$, and hence $C(u,v)$ throughout Fig.~\ref{fig:InitialCond_plates} will be different.
The theory is causal but \emph{nonlocal}, despite the locality of Eqs.~\eqref{eq:thetauu}--\eqref{eq:thetavv} as functionals of $C\,$;
it is $C$ itself that carries the nonlocal information.

Importantly, in Fig.~\ref{fig:InitialCond_plates}, plate $B$ is assumed to be at the same \emph{coordinate} distance $L$ from plate $A$ in both the hatted- and tilde-coordinate systems.
This is done without loss of generality, but one must be careful to scale the conformal factor and coordinates appropriately to make sure this is true.

While we will apply this theory to  a \emph{falling} Casimir apparatus
(i.e., geodesic motion of the center with a fixed separation between the plates),
this general theory applies to more arbitrary trajectories.

\subsection{Perturbation theory}\label{sec:perttheory}

To find an arbitrary solution, we appeal to perturbation theory.
In particular, the perturbation theory we consider is for $L$ small compared to both the curvature $R$ and the inverse acceleration of the plates.
We additionally  consider only the causal region in the future of region C in Fig.~\ref{fig:InitialCond_plates}; in this region, we solve Eq.~\eqref{eq:H_solve}.

In a sense that we now make precise, the Casimir apparatus will consist of two plates kept an equal distance $L$ from each other.
We assume the center of mass after $t=0$ (and into the future of region C) follows a timelike geodesic $P_0 = (U_0(\tau), V_0(\tau))$, where $\tau$ is defined as the proper time of this geodesic.
This free-fall trajectory satisfies the equations
\begin{equation}
  \begin{split}
  C(U_0(\tau),V_0(\tau)) U_0'(\tau) V_0'(\tau) & = 1, \\
  U_0''(\tau) + \Gamma\indices{^u_{uu}} U_0'(\tau)^2 & = 0, \\
  V_0''(\tau) + \Gamma\indices{^v_{vv}} V_0'(\tau)^2 & = 0.
\end{split}
\end{equation}
At a given time $\tau$, we can define a spacelike geodesic that connects the two plates;
this spacelike surface is orthogonal to $P_0(\tau)$ and parametrized by its proper distance $\eta$ so that
\begin{align}
  \partial_\eta^2 U_\eta(\tau) + \Gamma\indices{^u_{uu}} \partial_\eta U_\eta(\tau)^2 & = 0,  \label{eq:geodesic_etaU}\\
  \partial_\eta^2 V_\eta(\tau) + \Gamma\indices{^v_{vv}} \partial_\eta V_\eta(\tau)^2 & = 0,  \label{eq:geodesic_etaV}
\end{align}
with initial conditions
\begin{equation}
  \begin{split}
  (U_\eta(\tau), V_\eta(\tau))|_{\eta=0} & = (U_0(\tau),V_0(\tau)), \\
  (\partial_\eta U_\eta(\tau), \partial_\eta V_\eta(\tau))|_{\eta=0} & = (-U_0'(\tau),V_0'(\tau)).
\end{split}
\end{equation}
(The latter initial condition comes from the spacelike vector orthogonal to the trajectory's two-velocity.)
These coordinates $(\tau,\eta)$ are called Fermi coordinates~\cite{manasseFermiNormalCoordinates1963}, and in order for them to be defined everywhere between the plates (for all relevant times), we assume $L$ is sufficiently small.

Abusing our notation, we will use $A$ and $B$ as both coordinates and labels for the two plates, $A = -L/2$ and $B = L/2$.
We note that $L$ is now a \emph{physical}, not just a coordinate, distance; the plates are kept a fixed distance $L$ from each other (but do not themselves follow  geodesics).
In this sense, the center of mass of the Casimir apparatus is in free fall.

This setup allows us to perform a perturbation theory assuming $U_0'''(\tau) L^2 \ll U_0''(\tau) L \ll 1$ and $V_0'''(\tau) L^2\ll V_0''(\tau) L \ll 1$. (These conditions can alternatively be written in terms of Christoffel symbols and their derivatives.)
The details of that perturbation theory are in Appendix~\ref{sec:perturbation-theory}, but the result is that we perturbatively solve Eq.~\eqref{eq:H_solve} and Eq.~\eqref{eq:Hcircp_solve} under the above conditions to obtain
\begin{align}
  H'(v) & = \frac{1}{V_0'\circ V_0^{-1}(v)}\left[1 + \frac{R}{48}L^2 + O(L^4)  \right], \label{eq:Hpert} \\
  (H\circ p)'(u) &  = \frac{1}{U_0'\circ U_0^{-1}(u)}\left[1 + \frac{R}{48}L^2 + O(L^4)  \right], \label{eq:Hcircpert}
\end{align}
where $R$ is the Ricci scalar as defined by Eq.~\eqref{eq:RicciScalar}.

\subsection{Initial conditions}\label{sec:initcond}

With our perturbative solution \eqref{eq:Hpert}--\eqref{eq:Hcircpert}, we have a particular solution, but now we need to consider Fig.~\ref{fig:InitialCond_plates} in order to get a solution with initial conditions.
To aid in this task, we make an assumption on our metric: It has a timelike Killing vector defined by $\partial_t\,$; therefore, we assume
\begin{equation}
  C(u,v) = C(v - u)
\end{equation}
in the starting coordinates.
Further, in the past of Region C in Fig.~\ref{fig:InitialCond_plates}, the plates are on paths that follow the Killing vector field.
For ease of notation, we also define the remainder $r(v) \equiv \{ H(v) \}_{t_1-L}^{t_1 + L}\,$, and hence
\begin{equation}
    \tilde H(v) = 2 n L + \begin{cases}
  (H\circ p)^{-1}[r(v)], & r(v) < 0, \\
  H^{-1}[r(v)], & r(v) > 0.
  \end{cases}
\end{equation}
With this setup, we can match the perturbative solution of the previous section to the vacuum in the past of C, and we obtain a function $\tilde H$ (as described in Appendix~\ref{sec:initial_conditions}) given by
\begin{multline}
  \tilde H(v) = H(v) + \left(\frac12\Gamma_0 L + \frac1{6} [\Gamma_0 L]^2  \right) r(v) \\ - \mathrm{sgn}[r(v)] \frac12\left(1 + \Gamma_0 L  \right)\Gamma_0 \, r(v)^2 \\
  + \frac1{3} \Gamma_0^2 \, r(v)^3 + \cdots, \label{eq:Htilde_pert}
\end{multline}
where $\Gamma_0 \equiv \Gamma\indices{^v_{vv}}|_{v-u=x_0}$ and $x_0$ is the coordinate position that is a proper distance $L/2$ from either plate at the moment they are ``dropped'' (defined formally by Eq.~\eqref{eq:midwayx0}).

The more interesting object for our calculation of $\braket{T_{\mu \nu}}$ is the derivative of this function, which can be obtained from $\tilde H' = (Q' \circ H) H'$.
The function $Q(w) = w + \Omega(w)$ and the periodic function $\Omega$ can be read off from Eq.~\eqref{eq:Htilde_pert} as
\begin{equation}
  \Omega(w) = \frac1{6}  \Gamma_0 w(L - | w|)[3 + (L - 2|w|) \Gamma_0] + O(L^4),
\end{equation}
for $\quad w\in[-L,L]$.
Clearly this is continuous ($\Omega(L) = \Omega(-L) = 0$).
We also have a continuous first derivative:
\begin{equation}
  \begin{split}
  \Omega'(0^+)  = \Omega'(0^-) & = \frac16 L \Gamma_0( 3 + L \Gamma_0), \\
  \Omega'(L)  = \Omega'(-L) & = -\frac16 L \Gamma_0(3 - L \Gamma_0).
\end{split}
\end{equation}
But we start seeing discontinuities in the second derivative:
\begin{equation}
  \begin{split}
  \Omega''(0^+) & = - \Omega''(0^-) = - \Gamma_0( 1+ L \Gamma_0 ), \\
  \Omega''(L) & = - \Omega''(-L) = -\Gamma_0(1-L \Gamma_0 ).
\end{split}
\end{equation}
In fact, at this order, the third derivative has $\delta$-functions at $w = 0$ and $L$ amidst a constant background, while the fourth and higher are derivatives of the $\delta$-functions at these two points.

Therefore, we can write
\begin{multline}
  \tilde H'(v) = H'(v)\left[ 1 + \frac12 L \Gamma_0 + \frac1{6} \Gamma_0^2 L^2 \right.\\ \left. - ( 1 + \Gamma_0 L) \Gamma_0|r(v)| + \Gamma_0^2 r(v)^2\right].
\end{multline}
And if $w \in (-L,L]$, we have
\begin{multline}
  Q'(w) =  1 + \frac12\Gamma_0 L + \frac1{6} \Gamma_0^2 L^2  \\ - ( 1 + \Gamma_0 L) \Gamma_0|w| +  \Gamma_0^2 w^2 + \cdots.
\end{multline}
We can compute higher derivatives of $Q$ to aid in the calculation for the dynamical Casimir force,
\begin{equation}
  Q''(w) =  - ( 1 + \Gamma_0 L) \Gamma_0 \,\mathrm{sgn}(w) + 2\Gamma_0^2 w + \cdots
\end{equation}
and finally
\begin{multline}
  Q'''(w) = 2\Gamma_0^2 - 2 \Gamma_0 (1+\Gamma_0 L ) \delta(w) \\ + 2\Gamma_0(1-L \Gamma_0)\delta(w - L) + \cdots ,
\end{multline}
where in the last line the domain we calculate $Q'''(w)$ for is $(-L+\epsilon, L + \epsilon] $ for a small $\epsilon$.
These $\delta$-functions will become important with the dynamical Casimir effect: They resemble classical photons that bounce between the two plates that are in free fall (cf.\ \cite{fullingRadiationMovingMirror1976}).

\section{The Energy-Momentum Tensor}\label{sec:EnergyMomentumTensor}

In this section, we concentrate on calculating $\theta_{\mu\nu}\,$, and we dedicate separate sections to $\theta_{\mu\nu}^{\mathrm{stat}}$ and $\theta_{\mu\nu}^{\mathrm{dyn}}$ as well as to $\theta_{\mu\nu}$ inside and outside of the apparatus.

The values of this tensor as stated in Eqs.~\eqref{eq:thetauu} and \eqref{eq:thetavv} are found in the tilde coordinates $(\tilde u, \tilde v)$ that we previously calculated in Sec.~\ref{sec:GeneralTheory} with $\tilde u = \tilde H \circ p(u)$ and $\tilde v = \tilde H(v)$. The bulk of this section is spent transforming back to $(u,v)$ coordinates, so that we can make sense of observables  in Sec.~\ref{sec:CasimirEffect}.

\subsection{Static Casimir contribution}\label{StaticCasimirForce}

Apply the coordinate transformation to \eqref{zystat}:
\begin{equation}
  \begin{split}
  \theta^{\mathrm{stat}}_{uu} & = -\frac{\pi}{48L^2}[(\tilde H\circ p)'(u)]^2, \\
  \theta^{\mathrm{stat}}_{vv} & = -\frac{\pi}{48L^2}[\tilde H'(v)]^2.
\end{split}
\end{equation}
Substituting what we found previously, we have up to order $L^0$
\begin{widetext}
\begin{align}
 \theta_{uu}^{\mathrm{stat}} & = -\frac{\pi}{48[U_0'\circ U_0^{-1}(u)]^2 L^2} \left(1 + \frac1{24} R L^2 \right)\left(  1 + \Gamma_0 L +\frac7{12} \Gamma_0^2 L^2 - ( 2 + 3\Gamma_0 L) \Gamma_0|r\circ p(u)| + 3 \Gamma_0^2(r\circ p(u))^2 \right), \label{eq:thetauuStatic}\\
 \theta_{vv}^{\mathrm{stat}} & = -\frac{\pi}{48[V_0'\circ V_0^{-1}(v)]^2 L^2} \left(1 + \frac1{24} R L^2 \right)\left(  1 + \Gamma_0 L +\frac7{12} \Gamma_0^2 L^2 - ( 2 + 3\Gamma_0 L) \Gamma_0|r(v)| + 3 \Gamma_0^2[r(v)]^2 \right). \label{eq:thetavvStatic}
\end{align}
\end{widetext}

We see explicitly here dependence on $r(v)$, showing that this negative energy is behaving like a fluid in a box and sloshing back and forth.
To see the average effect on this Casimir apparatus, we define $\overline{f}(v) \equiv \frac1{2L}\int_0^{2L} f[v,r(v)] dr(v)$ as a semi-running average.
Applying this to the above amounts to averaging over a period of the periodic function $\Omega$, and we obtain
\begin{equation}
  \begin{split}
 \overline{\theta_{uu}^{\mathrm{stat}}} & =-\frac{\pi}{48[U_0'\circ U_0^{-1}(u)]^2 L^2} \left[1 + \frac1{24} (R+2 \Gamma_0^2) L^2 \right], \\
 \overline{\theta_{vv}^{\mathrm{stat}}} & =-\frac{\pi}{48[V_0'\circ V_0^{-1}(v)]^2 L^2} \left[1 + \frac1{24} (R+2 \Gamma_0^2) L^2 \right].
\end{split}
\end{equation}

The first thing we notice is the dependence on the null coordinates $(u,v)$.
A free-fall observer located at the midway point between the plates will observe the tensor with her timelike vector $(U_0'(\tau),V_0'(\tau))$.
Therefore, we can start to see the equivalence principle at work: the lowest-order term is exactly what this observer would expect from plates in flat space.
The next order term is $O(1)$ in 1+1D and captures both curvature and initial-condition corrections.
The curvature term is the beginning of what will be the tidal Casimir effect, as will become more apparent when we find the energy density as measured by comoving observers.
The initial-condition contributions give us non-trivial dependence on $(u,v)$ and an overall shift to the energy as captured by these averaged quantities; while these represent excitations between the plates, it is in fact lowering the already negative energy.

\subsection{Dynamic contribution}\label{sec:dynamicContribution}

Using properties of the $F_x(f)$, we first note that
\begin{align}
  \theta_{uu}^{\mathrm{dyn}} & =\frac1{24\pi}\left[  F_u(C)  - F_u[(\tilde H\circ p)']\right], \label{eq:thetatdynuu}\\
  \theta_{vv}^{\mathrm{dyn}} & =\frac1{24\pi}\left[  F_v(C) - F_v(\tilde H')  \right]. \label{eq:thetatdyn}
\end{align}
The first terms [$F_u(C)$ and $F_v(C)$] were previously evaluated in Eqs.~\eqref{eq:FuChristofell} and \eqref{eq:FvChristofell}.
The second term in Eq.~\eqref{eq:thetatdyn} is
\begin{equation}
  F_v[\tilde H'] = F_v[(Q'\circ H)H'] = H'(v)^2 F_w[Q'] + F_v[H'],
\end{equation}
where $w = H(v)$.
We are only interested in $L^0$ terms, so we can calculate
\begin{equation}
  F_w[Q'] 
   = \frac12 \Gamma_0^2 - 2 \Gamma_0 \sum_n\left[\delta(w + 2 n L) - \delta(w + 2nL-L)\right],
\end{equation}
and, again to order $L^0$,
\begin{equation}
  \begin{split}
  F_v[H'] & = F_v\left[\frac1{V_0'\circ V_0^{-1}}\right]\\ & = - \frac1{[V_0'\circ V_0^{-1}(v)]^2} F_{V_0^{-1}(v)}[V_0'],
\end{split}
\end{equation}
Since the center of the apparatus follows a geodesic, $F_{\tau}[V_0']$ is easily calculated using the geodesic equation and its derivative [see Eq.~\eqref{eq:geodesic-derivative}].
This yields
\begin{equation}
  \frac1{(V'_0)^2}F_\tau[V_0'] = - \left[\partial_v \Gamma\indices{^v_{vv}} - \tfrac12 (\Gamma\indices{^v_{vv}})^2 \right] +\frac14 \frac{R}{(V_0')^2}\,.
\end{equation}
Therefore,
\begin{equation}
  F_v[H'] = \left[\partial_v \Gamma\indices{^v_{vv}} - \tfrac12 (\Gamma\indices{^v_{vv}})^2 \right] - \frac14 \frac{R}{(V_0'\circ V_0^{-1}(v))^2}\,.
\end{equation}
The Christoffel symbols in this expression are evaluated along the geodesic path at proper time given by $V_0^{-1}(v)$.
To the order at which we are evaluating things currently, the terms that go as the Christoffel symbols from both $F_v[C]$ and $F_v[H']$ cancel one another (corrections being of order $L^1$ and hence omitted from our equations).

Similar calculations can be conducted for Eq.~\eqref{eq:thetatdynuu}; the final formulas are then
\begin{widetext}
\begin{align}
  \theta_{uu}^{\mathrm{dyn}} & = -\frac1{24\pi} \frac1{U_0'\circ U_0^{-1}(u)^2}\left\{\frac12 \Gamma_0^2 - 2 \Gamma_0 \delta_{2L}[H\circ p(u)] + 2\Gamma_0 \delta_{2L}[H\circ p(u)-L] -\frac R 4 \right\}, \label{eq:thetauuDyn} \\
  \theta_{vv}^{\mathrm{dyn}} & = -\frac1{24\pi}\frac1{V_0'\circ V_0^{-1}(v)^2}\left\{\frac12 \Gamma_0^2 - 2 \Gamma_0 \delta_{2L}[H(v)] + 2\Gamma_0 \delta_{2L}[H(v)-L] -\frac R 4 \right\}, \label{eq:thetavvDyn}
\end{align}
where $\delta_{2L}(x)$ is a Dirac ``comb''\negthinspace, $\delta_{2L}(x) \equiv \sum_n \delta(x - 2 n L)$.
These terms represent null trajectories that bounce back and forth between the plates, highlighting that we have excitations between the plates as they fall.
\end{widetext}

We can again average the result much as before:
\begin{equation}
  \begin{split}
  \overline{\theta_{uu}^{\mathrm{dyn}}} & = -\,\frac1{96\pi}\frac{2\Gamma_0^2 -R }{U_0'\circ U_0^{-1}(u)^2}\, , \\
  \overline{\theta_{vv}^{\mathrm{dyn}}} & = -\,\frac1{96\pi}\frac{2 \Gamma_0^2 -R }{V_0'\circ V_0^{-1}(v)^2}\,.
\end{split}
\end{equation}

In these expressions, we see that we get a nonzero contribution to the energy-momentum tensor only if $\Gamma_0$ or $R$ is nonzero --- this term is the generalization of the dynamical Casimir effect in flat space.
In curved space, we notice that independent of our initial state we have an effect proportional to $R$ that can still be interpreted in the context of the dynamical Casimir effect: Falling plates experience a classical tidal force, but we have imposed that they retain a fixed proper separation.
This means that some outside force (e.g., a rod of fixed length) is keeping the plates on course;
the resulting acceleration  creates a dynamical response in the energy-momentum tensor represented by $R$.
On the other hand, the presence of $\Gamma_0$ in this expression is more straightforward: The plates are initially at rest, we suddenly begin moving them, and that creates a response in the energy-momentum tensor here just as it does in flat space.

\subsection{Outside the Casimir apparatus}\label{sec:OutsideApparatus}

To obtain the force on the plates from the energy-momentum tensor, it is necessary to include the outside of the plates.
Different vector fields may define the Hamiltonian to the left and right of the plates, and we encode that information by using different coordinates in the two regions: in the region left of A,  $(u^<, v^<)$ with conformal factor $C^<$, and in the region left of B, $(u^>, v^>)$ with conformal factor $C^>$.
The two corresponding vector fields  are $\partial_{t^<}$ and $\partial_{t^>}\,$.
To the right of plate $B$ (see Fig.~\ref{fig:coordinate-transformations}), we only need to go to coordinates $(\bar u^>, \bar v^>) = (p^>(u^>),v^>)$ to get the appropriate causal structure, and in that case we  have only a dynamical part to worry with and we define it as
(dropping the superscript $>$ on coordinates for ease of reading)
\begin{equation}
  \theta_{uu}^{>} = \frac1{24\pi}\left[F_u(C^>)-  F_u(p^{>\prime}) \right].
\end{equation}
Recall that $p^> = V^>_A\circ U_A^{>-1}$, and to lowest order we can take $A=0$.
Then
\begin{equation*}
p^{>\prime}(u) = \frac{V_A^{>\prime}[U_A^{>-1}(u)]}{U_A^{>\prime}[U_A^{>-1}(u)]}
\end{equation*}
and
\begin{equation}
  \begin{split}
  F_u(p^{>\prime}) & = \frac{F_{U^{>-1}_0(u)}[V^{>\prime}] - F_{U_0^{>-1}(u)}(U_0^{>\prime})}{U^{>\prime}_0[U_0^{>-1}(u)]^2}  \\
    & = -\frac{(V^{>\prime}_0)^2}{(U^{>\prime}_0)^2}\left[\partial_v \Gamma\indices{^{>v}_{vv}} -\frac12(\Gamma\indices{^{>v}_{vv}})^2\right] \nonumber \\ & \phantom{===} +\partial_u \Gamma\indices{^{>u}_{uu}} - \frac12(\Gamma\indices{^{>u}_{uu}})^2.
  \end{split}
\end{equation}
Also, on this side of the plates we have $\theta_{vv}^{>} = \frac1{24\pi} F_v(C^>)$.
As always, we need to be careful about where we evaluate the Christoffel symbols, but for the force calculations we are interested in, we will be evaluating them on the geodesic paths where $F_u(C)$ matches the appropriate part of $F_u(p')$.
Thus we have near plate $B$
\begin{equation}
  \begin{split}
  \theta_{uu}^> & = -\frac1{24\pi}\frac{(V^{>\prime}_0)^2}{(U^{>\prime}_0)^2}\left[\partial_v \Gamma\indices{^{>v}_{vv}} -\frac12(\Gamma\indices{^{>v}_{vv}})^2\right], \\
  \theta_{vv}^> & = -\frac1{24\pi}\left[\partial_v \Gamma\indices{^{>v}_{vv}} -\frac12(\Gamma\indices{^{>v}_{vv}})^2\right].
\end{split}
\end{equation}

A similar argument can be made for the region left of A, but the roles of $u$ and $v$ are opposite.
The result is that near plate $A$,
\begin{equation}
  \begin{split}
  \theta_{uu}^< & = -\frac1{24\pi}\left[\partial_u \Gamma\indices{^{<u}_{uu}} -\frac12(\Gamma\indices{^{<u}_{uu}})^2\right], \\
  \theta_{vv}^< & = -\frac1{24\pi}\frac{(U^{<\prime}_0)^2}{(V^{<\prime}_0)^2}\left[\partial_u \Gamma\indices{^{<u}_{uu}} -\frac12(\Gamma\indices{^{<u}_{uu}})^2\right].
\end{split}
\end{equation}

We again stress that in general $\Gamma\indices{^{<u}_{uu}}$, $\Gamma\indices{^u_{uu}}$, and $\Gamma\indices{^{>u}_{uu}}$ are not equal to each other, because the vacuum states are defined by  different vector fields (or conformal factors $C$).
Ordinarily we have in mind an initially static configuration, so that the states are uniquely determined,
but the motion of a plate will induce a conformal mapping of the $f(u)$ type on one side and the $g(v)$ type on the other, producing quite different results for $C$.
In particular, outside of a black hole one can choose different vacua (Hawking-Hartle/Boulware/Unruh), and we want to reserve the freedom to change the vacuum on either side of the plate.

\section{Casimir effect}\label{sec:CasimirEffect}

\subsection{The Casimir force}\label{sec:CasimirForceSubsec}

 To calculate the Casimir force, we have to tease out what is happening between the plates versus outside  the plates.
Plate $B$, for instance, in its reference frame experiences a pressure coming from  the different tensors $T_{\mu\nu}$ on either side of it.

The pressure in the energy-momentum tensor is given by the purely spatial component, but we need to be careful how to specify this for the plate.
The plate is defined as remaining a fixed distance from a geodesic, and so its 2-velocity is orthogonal to the spacelike vector $(\partial_\eta U_\eta(\tau), \partial_\eta V_\eta(\tau))$ with $\eta=L/2$.
In other words, the 2-velocity is $(-\partial_\eta U_\eta(\tau), \partial_{\eta} V_\eta(\tau))$ (again, with $\eta = L/2$).
Therefore, we need to consider $\braket{T_{\eta \eta}}_B$,
\begin{multline}
  \braket{T_{\eta\eta}} = \braket{T_{uu}}(\partial_\eta U_\eta(\tau))^2 + \braket{T_{vv}}(\partial_\eta V_\eta(\tau))^2 \\ + 2 \braket{T_{uv}}\partial_\eta U_\eta(\tau) \partial_\eta V_\eta(\tau) \label{eq:Tetaeta}
\end{multline}
This quantity is different on the two sides of the plate, yielding a net force on the plate.
Indeed, that force  is given by
\begin{equation}
  F_B = \braket{T_{\eta \eta}}_{B^-} - \braket{T_{\eta \eta}}_{B^+},
\end{equation}
the pressure from the left of the plate minus the pressure from the right.
In this quantity, parts of $\braket{T_{\mu\nu}}$ that are the same on both sides of the plates  (such as the term proportional to $R g_{\mu\nu}$) will cancel and  will therefore be neglected in what follows.
To the order we have worked ($L^0$ in $\braket{T_{\mu\nu}}$), we can find the pressure due to the dynamical effect to the right of plate $B$ by
\begin{equation}
  \begin{split}
  \braket{T_{\eta\eta}}_{B^+} & \equiv U_0'(\tau)^2 \theta^>_{uu} + V_0'(\tau)^2\theta^>_{vv}, \\
  & = \frac{[V_0^{>\prime}]^2}{12\pi} \left[ \partial_v \Gamma\indices{^{>v}_{vv}} -\frac12(\Gamma\indices{^{>v}_{vv}})^2\right].
\end{split}
\end{equation}

 Between the plates we can combine the effects of $\theta^{\mathrm{dyn}}_{\mu\nu}$ and $\theta^{\mathrm{stat}}_{\mu\nu}$.
For the static contribution we cannot just use the $(U_0(\tau),V_0(\tau))$ trajectory because $\theta^{\mathrm{stat}}_{\mu\nu}$ is of order $L^{-2}$, so we need to consider $(\partial_\eta U_\eta(\tau), \partial_\eta V_\eta(\tau))$.
From $\theta_{uu}^{\mathrm{stat}} [\partial_\eta U_\eta(\tau)]^2 + \theta_{vv}^{\mathrm{stat}} [\partial_\eta V_\eta(\tau)]^2$, we see that the terms that go as $L^{-2}$ are multiplied by
\begin{align}
\frac{\partial_\eta U_\eta(\tau) }{U_0'[U_0^{-1}(U_{\eta}(\tau))]} & =  1 - \frac1{4} R \eta^2 + \cdots, \label{eq:pdetaU} \\
\frac{\partial_\eta V_\eta(\tau) }{V_0'[V_0^{-1}(U_{\eta}(\tau))]} & =  1 - \frac1{4} R \eta^2 + \cdots. \label{eq:pdetaV}
\end{align}
Therefore, with the plates at $\eta = L/2$, we have
\begin{multline}
    \overline{\braket{T_{\eta\eta}}}_{B^-} =  -\frac{\pi}{24 L^2} \\ \times \left[ 1  - \frac1{12} (R - \Gamma_0^2) L^2 - \frac{1}{2\pi^2}(R- 2 \Gamma_0^2) L^2\right],
\end{multline}

The total contribution to the Casimir force as experienced by the plate is then
\begin{widetext}
\begin{equation}
  \overline{F}_B  = -\frac{\pi}{24 L^2} + \frac{1}{48}\left(\frac{\pi}{6}+\frac1{\pi} \right)R - \frac1{24}\left( \frac{\pi}{12} - \frac1{\pi}\right)\Gamma_0^2 - \frac{(V_0^{>\prime})^2}{12\pi} \left[\partial_v \Gamma\indices{^{>v}_{vv}} -\tfrac12 (\Gamma\indices{^{>v}_{vv}})^2 \right] + O(L).\label{eq:ForcePlateB}
\end{equation}
Similarly, we can calculate the force on plate $A$, defined as $F_A = \braket{T_{XX}}_{A^+} - \braket{T_{XX}}_{A^-}$ so that a negative force implies attraction to the other plate:
\begin{equation}
 \overline{F}_A  = -\frac{\pi}{24 L^2} + \frac{1}{48}\left(\frac{\pi}{6}+\frac1{\pi} \right)R - \frac1{24}\left(\frac{\pi}{12} - \frac1{\pi} \right)\Gamma_0^2 - \frac{(U_0^{<\prime})^2}{12\pi} \left[\partial_u \Gamma\indices{^{<u}_{uu}} -\tfrac12 (\Gamma\indices{^{<u}_{uu}})^2 \right] + O(L).\label{eq:ForcePlateA}
\end{equation}
\end{widetext}
Recall that (unlike in our earlier, more general considerations) $L$ is now normalized to be the physical distance between the plates.
Eqs.~\eqref{eq:ForcePlateB} and \eqref{eq:ForcePlateA} describe the forces experienced by the falling plates.
At lowest order ($L^{-2}$), the plate experiences the Casimir force in the normal way; by the equivalence principle the lowest order term should be that of flat space.
But at next order ($L^0$) both curvature and initial conditions start to affect the force.
The initial conditions (represented by $\Gamma_0$) seem to increase the attractive Casimir force somewhat counterintuitively: The process of dropping the plates has created excitations between the plates which, instead of pushing the plates away from each other, are pulling the plates together. 
The curvature can increase or decrease the Casimir force depending on its sign.
In particular the term that goes as $\frac{\pi}{288} R$ is what we call the \emph{tidal Casimir effect}, and it is not captured by taking the derivative of the total Casimir energy with respect to $L$, as explained in the next section.
Finally, the final term is the radiation pressure of the falling plate, which would be present even without two Casimir plates.
Indeed, one can find the force on a single plate by calculating $F_B - F_A$.
However, the Casimir apparatus splits this force between the two plates, as one would expect.

\subsection{The static Casimir energy}\label{sec:staticCasimirEnergy}

In this section we drop the generality of the previous sections to concentrate on the static Casimir energy.
First, we assume $\Gamma_0 = 0$ (so that there are no dynamical terms).
We can write the general metric in terms of Fermi coordinates $(\tau,\eta)$, and we find up to order $L^2$
\begin{equation}
  ds^2 = \left(1 + \frac12 R(\tau) \eta^2 \right)d\tau^2 - d\eta^2,
\end{equation}
where $\eta\in[-L/2,L/2]$.
If we assume $R$ changes slowly compared to $\eta$, then this expression has an approximate Killing vector $\partial_\tau$, and we can define an approximately conserved energy
\begin{equation}
    E = \int \braket{T\indices{^{\tau}_{\tau}}} \sqrt{-g}\, d^2 x. \label{eq:energy-conservation}
\end{equation}
With this, we can isolate an energy density
\begin{equation}
 \rho(\eta) = \braket{T\indices{^{\tau}_\tau} } = \braket{T_{\tau \tau}}\left( 1 - \frac12 R \eta^2 + \cdots\right). \label{eq:CasimirEnergyDensity}
\end{equation}
For this simple situation (specifically when $R$ is $\tau$-independent), we can short-circuit the previous analysis to obtain the Casimir result by defining $\tilde \eta = \eta - \frac1{12} R \eta^3 + \cdots$ so that $d\eta = \left(1 + \frac12 R \eta^2 + \cdots \right)^{1/2} d\tilde\eta$ using $R \eta^2\ll 1$.
These $(\tau,\tilde\eta)$ coordinates are conformally flat and using Eqs.~\eqref{eq:thetauu} and \eqref{eq:thetavv}, we find
\begin{equation}
  \braket{T^{\mathrm{cas}}_{\tau\tau}} = -\frac{\pi}{24 L^2}\left( 1 + \frac1{24} R L^2  + \cdots \right).
\end{equation}
Therefore, we have the energy density
\begin{equation}
  \rho_{\mathrm{cas}}(\eta) = - \frac{\pi}{24 L^2} \left(1 + \frac1{24}R L^2 - \frac12 R \eta^2 + \cdots \right). \label{eq:cas-energydensity}
\end{equation}
We can integrate $\rho_{\mathrm{cas}}(\eta)$ according to Eq.~\eqref{eq:energy-conservation} to obtain
\begin{equation}
  E_{\mathrm{cas}} = - \frac{\pi}{24 L}\left(1 + \frac1{48} R L^2 + \cdots \right). \label{eq:CasimirEnergy}
\end{equation}
Before proceeding, notice that $ - \partial E_{\mathrm{cas}}/\partial L$ does not reproduce the force we expect.
We crucially obtain the wrong numerical coefficient for what we dubbed the ``tidal Casimir force'': the term that went as $\frac{\pi}{288}R$ 
 \footnote{We point out, however, that $\rho_{\mathrm{cas}}(L/2) = -\tfrac{\pi}{24L^2} + \tfrac{\pi}{288}R + \cdots $ has the term that we labeled as ``tidal'' in the force calculation ($\tfrac{\pi}{288} R$).}.
However, notice that the energy density is tidally spread out over $\eta$ as indicated by Eq.~\eqref{eq:CasimirEnergyDensity}.
Our interpretation is that the force is related to the local energy density at the plate, rather than the total energy in the apparatus.

To make this precise, the force can be derived by considering the total Killing energy in the system (defined by the Killing vector field $\partial_\tau$).
In addition to the Casimir energy, this includes the energy-momentum vector of the plates.
If plate $B$ is of mass $m$, then its four momentum is
\begin{equation}
  p^\mu = m u^\mu, \quad u^{\mu} = \left[ \sqrt{1 + \dot \eta_B^2}(1 - \tfrac14 R \eta_B^2 + \cdots) ,\dot \eta_B \right],
\end{equation}
where $\eta_B$ is the position of plate $B$, and $\dot \eta_B = \frac{d\eta_B}{ds}$ where $s$ is the proper time of the plate.
Then, the conserved quantity is
\begin{equation}
  E = p_\tau + E_{\mathrm{Cas}},
\end{equation}
which is (up to order $RL^2$)
\begin{multline}
  E = -\frac{\pi}{24 (\eta_B + L/2)} - \frac{\pi}{1152}R(L/2 + \eta_B) \\ + m  \sqrt{1 + \dot \eta_B^2}(1 +\tfrac14 R \eta_B^2).
\end{multline}
We can obtain a force equation by taking a derivative with respect to $s$, and we find
\begin{multline}
  m \ddot \eta_B = - \frac{\pi \sqrt{1 + \dot \eta_B^2}(1 - \frac1{48} R L^2 - \frac14 R \eta_B^2)}{24(\eta_B + L/2)} \\ - \frac12 m (1 + \dot \eta_B^2)R \eta_B.
\end{multline}
The second term is a gravitational force (and can be derived from the geodesic equation), and we can find the Casimir force by letting $\dot\eta_B=0$ and $\eta=L/2$, so that the force follows:
\begin{equation}
  m\ddot \eta_B = -\frac{\pi}{24 L^2} + \frac{\pi}{288} R - \frac14 m RL. \label{eq:newtonequation}
\end{equation}
Therefore, the static Casimir force is
\begin{equation}
   F_{\mathrm{cas}} = -\frac{\pi}{24 L^2} + \frac{\pi}{288} R + \cdots. \label{eq:CasimirForceStatic}
\end{equation}
This agrees with Eqs.~\eqref{eq:ForcePlateA} and \eqref{eq:ForcePlateB} if we neglect all dynamical effects.
While the previous results are more general, the agreement found here shows how $\braket{T_{\eta\eta}}$ is related to the force calculated from the existence of a time-like Killing direction.
And finally, note that $F_{\mathrm{cas}} \neq -\partial E_{\mathrm{cas}} / \partial L$, but $F_{\mathrm{cas}}$ has here been derived from the Newton's equation Eq.~\eqref{eq:newtonequation}.

Recall that to keep the plates at a fixed distance, there should be a rod between the plates balancing both Casimir and tidal forces, $F_{\mathrm{cas}}$ and $-\frac14 m RL$.

We end this section by returning to the full, time-dependent solution derived in previous sections.
First, any comoving observer at position $\eta$ between the plates ought to be able to measure the energy density.
The observer's worldline is given by $(U_\eta(\tau), V_\eta(\tau))$ and due to parallel transport, the two-velocity is given by $(-\partial_\eta U_\eta(\tau),\partial_\eta V_\eta(\tau))$.
Calling the observer's proper time $\tilde\tau$ (distinct from the center of mass's proper time, $\tau$), we can compute the measured energy density
\begin{multline}
  \braket{T_{\tilde \tau\tilde \tau}} = \braket{T_{uu}}(\partial_\eta U_\eta(\tau))^2 + \braket{T_{vv}}(\partial_\eta V_\eta(\tau))^2 \\ - 2 \braket{T_{uv}}\partial_\eta U_\eta(\tau) \partial_\eta V_\eta(\tau).
\end{multline}
This quantity is nearly identical to $\braket{T_{\eta\eta}}$ in Eq.~\eqref{eq:Tetaeta} except that the off-diagonal term $\braket{T_{uv}}$ contributes with opposite sign;  this term is purely determined by the curvature as seen in Eq.~\eqref{eq:EnergyMomentumTensor}.
In fact,
\begin{equation}
  - 2 \braket{T_{uv}}\partial_\eta U_\eta(\tau) \partial_\eta V_\eta(\tau)  = -\frac1{24\pi} R.
\end{equation}
The other terms can be found from Eqs.~\eqref{eq:thetauuStatic}, \eqref{eq:thetavvStatic}, \eqref{eq:thetauuDyn}, and \eqref{eq:thetavvDyn} so that we have
\begin{equation}
  \braket{T_{\tilde\tau\tilde \tau}} = \theta_{uu}[\partial_\eta U_\eta(\tau)]^2 + \theta_{vv} [\partial_\eta V_\eta(\tau)]^2-\tfrac1{24\pi} R. \label{eq:comoving-obs-energy-dens}
\end{equation}
This has a lot of out-of-equilibrium structure inherited from $\theta_{uu}$ and $\theta_{vv}$ which we will explore in the next section.
Here though, we set $\Gamma_0=0$ and we obtain for the static Casimir contribution
\begin{equation}
    \braket{T_{\tilde\tau\tilde \tau}^{\mathrm{cas}}} = \rho_{\mathrm{cas}}(\eta).
\end{equation}
In fact, a comoving observer at position $\eta$ will measure precisely the conserved energy density we previously derived in the more restrictive case where $R$ is $\tau$-independent, Eq.~\eqref{eq:cas-energydensity}.
And as previously observed, the static Casimir contribution is tidally spread out between the two plates---more negative energy has built up on the plates and the attractive force between the plates has increased.

\subsection{Out-of-equilibrium Casimir energy}\label{sec:outofequilibriumCasimir}

In the previous section, we saw that the Casimir energy gets tidally distributed between the plates, and the Casimir force increases as a result.
In this section, we see that this analogy with a fluid holds even in the out-of-equilibrium nature of that energy density.
Indeed, we can see how that energy density sloshes around between the plates.

\begin{figure}
  \includegraphics[width=\columnwidth]{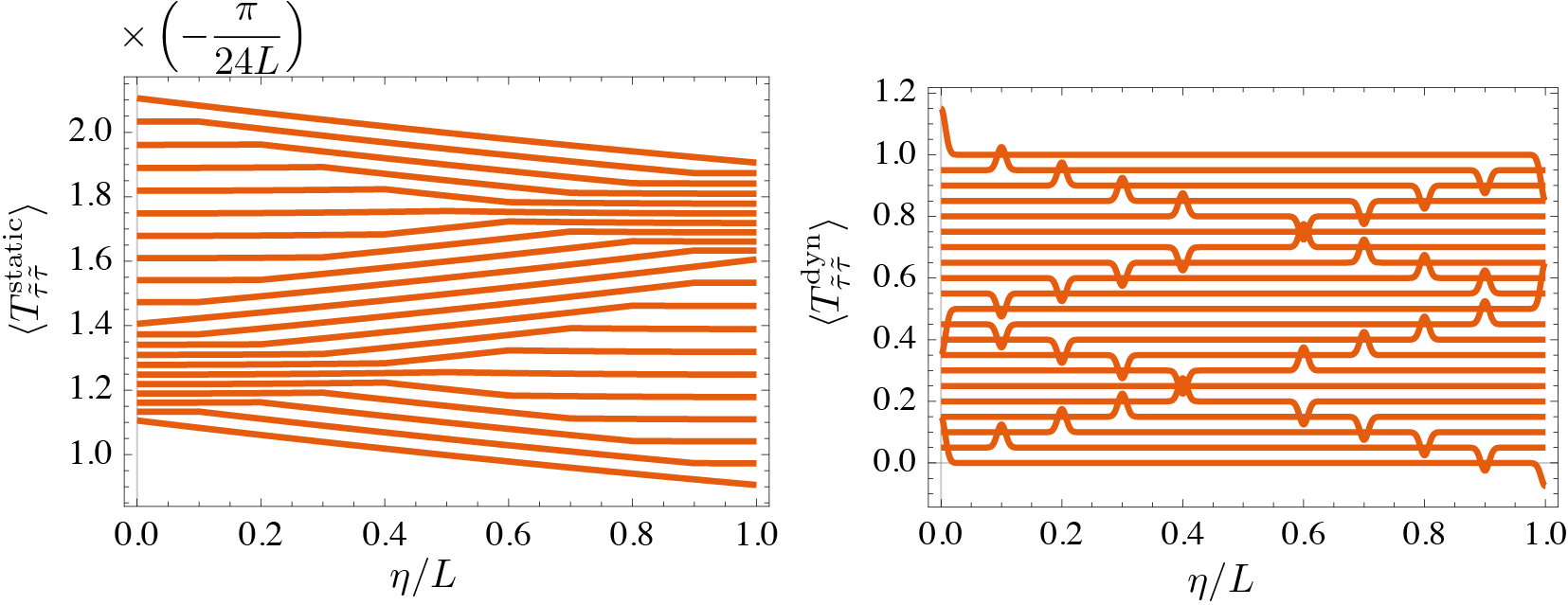}
  \caption{Plots of the static and dynamic contributions to $\braket{T_{\mu\nu}}$ [progression in time is shifted vertically by $+\tau/2$].
  (left) The Casimir energy is spread out in space and once it begins falling, it starts to slosh back and forth between the plates.
  (right) The dynamic Casimir effect creates small energy bursts at the point where the plates are dropped, which then bounce back and forth between the plates.}
  \label{fig:SloshPlot}
\end{figure}

If we return to Eq.~\eqref{eq:comoving-obs-energy-dens}, we can highlight what is occurring out of equilibrium, by letting $R=0$ and keeping $\Gamma_0$ is finite
(e.g.\ the Rindler coordinates with $C(v-u) \propto e^{\Gamma_0(v-u)}$ \cite{daviesEnergymomentumTensorEvaporating1976}, where $\Gamma_0$ describes the acceleration in a frame that is constantly accelerating).
The calculations can be done exactly (but match the approximations made earlier done for finite $R$ and $\Gamma_0$); the resulting
 components $\theta^{\mathrm{stat}}_{\mu\nu}$ and $\theta^{\mathrm{dyn}}_{\mu\nu}$  are pictured in Fig.~\ref{fig:SloshPlot}.
Since $\Gamma_0$ represents the initial conditions in our system, we find two interesting effects: first, $\theta^{\mathrm{stat}}_{\tilde \tau \tilde \tau}$ is sloshing back and forth between the plates with a period of $2L$, and second, from $\theta^{\mathrm{dyn}}_{\tilde \tau \tilde \tau}$, little packets of energy (positive and negative) are bouncing back and forth between the plates --- also with period $2L$.
In this model these packets are created solely by the nonuniform acceleration at the time when the plates are dropped.

The fluid analogy holds well for the static contribution: a container of fluid that is uniformly accelerated for a time until it is ``launched'' into an inertial frame where it will begin to slosh back and forth in the container.
As expected, the dynamical part breaks the analogy due to the excitations being created from the vacuum.
Nonetheless, these contributions can be neatly separated from each other.

\section{Falling into a black hole}\label{sec:blackhole}

We can apply all of our previous analysis to the problem of two plates falling into a black hole.  We assume the plates are dropped into the black hole from a Schwarzschild radial coordinate $r=r_0\,$.

First we need to determine the Cauchy surfaces with which our vacuum state is defined inside and outside of the plates.
Inside the plates, we take the initial vacuum as defined by the Killing field $\partial_t$ by our analysis in Appendix~\ref{sec:initial_conditions}.
Outside the plates, we retain more freedom to choose our initial vacuum, and that freedom is characterized by explicit dependence on Christoffel symbols in Eq.~\eqref{eq:ForcePlateA} and \eqref{eq:ForcePlateB}.

First, let us ignore the vacuum outside  the plates.
The metric between the plates is \cite{daviesEnergymomentumTensorEvaporating1976}
\begin{equation}
  ds^2 = \left( 1 - \frac{2GM}r \right) [dt^2 - (dr^*)^2],
\end{equation}
where $r^*  = r + 2 GM \log\left| \tfrac{r}{2GM}-1\right|$.
The null coordinates are defined by
\begin{equation*}
u = \frac{t - r^*}{\sqrt{1-2GM/r_0}}, \quad v =\frac{t + r^*}{\sqrt{1-2GM/r_0}}.
\end{equation*}
The curvature term (which will lead to a tidal Casimir effect)
is
\begin{equation}
 R = -\, \frac{4 GM}{r^3}\,,
\end{equation}
so if we look at the pressure on plate $B$ in Eq.~\eqref{eq:CasimirForceStatic} due to the curvature, the inward force between the plates and the Casimir energy \eqref{eq:CasimirEnergy} increase as the plates fall into the black hole.
However, if we look at Eq.~\eqref{eq:cas-energydensity} the energy density near the plate has \emph{decreased} (i.e.\ the magnitude of the negative energy density increased), and in fact that tidal value is  $\frac{\pi}{288} R < 0$ as explained in Section \ref{sec:staticCasimirEnergy}.
The negative energy is experiencing an extra tidal force and moves to the sides of the Casimir cavity;  as a result, the plates feel a stronger attractive force.

Additionally, the initial conditions can be used to find
\begin{equation}
  \Gamma_0  =  -\frac{GM}{r_0^2}\frac1{\sqrt{1 - 2GM/r_0}}\,.
\end{equation}
When $r_0\rightarrow \infty$ (a fall from infinity), one sees that there is no effect of initial conditions (this is, in some sense, an ``adiabatic'' limit).
Further, $\Gamma_0\rightarrow\infty$ at the horizon, indicating that dropping the plates from a stationary state near the horizon creates a burst of energy (and our perturbation theory breaks down quite severely).

If we let $r_0\rightarrow\infty$, we have the modified Casimir force
\begin{equation}
  F_{A,B} = -\frac{\pi}{24 L^2} - \frac{GM}{12 r^3} \left( \frac{\pi}6 - \frac1{\pi}\right) + F_{A,B}^{\mathrm{dyn}}\,,
\end{equation}
where $F_{A,B}^{\mathrm{dyn}}$ are the forces on plates A and B from the radiation pressure outside of the plates (caused by the dynamical Casimir effect).
On the other hand, the total energy between the plates is
\begin{equation}
  E_{\mathrm{cas}} = -\frac{\pi}{24 L} + \frac{GM}{288\pi r^3}L.
\end{equation}

To fully determine force, though, we need to consider the radiation pressure from falling, given by $F_{A,B}^{\mathrm{dyn}}\,$.
The objects $F_{A,B}^{\mathrm{dyn}}$ would be present with only one plate; in fact, for one plate we would have the radiation pressure from both sides so that $F_{\text{one plate}} = F_A^{\mathrm{dyn}} - F_B^{\mathrm{dyn}}$.
The choice of initial vacuum is important since we need both the Christoffel symbols and the trajectory in those coordinates.

On either side of the plates, we consider three states: the Hartle--Hawking vacuum, the Boulware vacuum, and the Unruh vacuum \cite{birrellQuantumFieldsCurved1984}.
The Hartle--Hawking and Boulware vacua are associated, respectively,  with the timelike Kruskal and Schwarzschild vector fields outside the black hole.
For our purposes, let us call the $(u,v)$ coordinates Schwarzschild and $(\underline{u},\underline{v})$ Kruskal coordinates.
The conversion between the two is
\begin{equation}
  \underline{u} = - e^{- u/(4GM)}, \quad \underline{v} = e^{v/(4GM)}, \label{eq:coord-trans-schwarz-krusk}
\end{equation}
so that
\begin{equation}
  \underline u \underline v = \left(1 - \tfrac{r}{2GM}\right)e^{r/2GM},
\end{equation}
and the metric for Kruskal coordinates is
\begin{equation}
  ds^2 =  \frac{4(2GM)^3}r e^{-r/2GM} d \underline u d\underline v.
\end{equation}
With these facts we can compute, in Kruskal coordinates for plate $A$ and trajectory $(\underline{U}_0(\tau),\underline{V}_0(\tau))$ (again, dropping the underline on coordinates in favor of one on the symbol)
\begin{multline}
  \partial_{u}\underline{\Gamma}\indices{^{u}_{uu}} - \tfrac12 (\underline{\Gamma}\indices{^{u}_{uu}} )^2 \\ = \frac{1}{\underline{u}^2}\left(\frac{2GM}{r} - 1\right)^2\left[ 1+ \frac{4 GM} r + \frac{3(2GM)^2}{r^2} \right].
\end{multline}
Isolating the dynamical part outside of the black apparatus from Eq.~\eqref{eq:ForcePlateA}, we have
\begin{multline}
    F_A^{\mathrm{dyn}}|_{\mathrm{HH}} = - \frac1{12\pi} \left[\frac{\underline{U}_0'(\tau)}{\underline{U}_0(\tau)}\right]^2 \left(\frac{2GM}{r} - 1\right)^2 \\ \times \left[ 1+ \frac{4 GM} r + \frac{3(2GM)^2}{r^2} \right],
\end{multline}
where HH indicates the Hartle-Hawking vacuum.
On the other hand, in Schwarzschild coordinates the trajectory is $(U_0(\tau),V_0(\tau))$ and we have
\begin{equation}
  \partial_{u}\Gamma\indices{^{u}_{uu}} - \tfrac12 (\Gamma\indices{^{u}_{uu}} )^2 = - \frac{2 G M}{2 r^3} \left( 1- \frac34\frac{2GM}{r} \right),
\end{equation}
and hence
\begin{equation}
  F_A^{\mathrm{dyn}}|_{\mathrm{B}}  =  \frac{U_0'(\tau)^2 (2 G M)}{24\pi r^3} \left( 1- \frac34\frac{2GM}{r} \right),
\end{equation}
where B indicates the Boulware vacuum.
Similar calculations can be done for $F_B^{\mathrm{dyn}}$ and it just amounts to letting $\underline{U}_0 \rightarrow \underline{V}_0$ and ${U}_0 \rightarrow {V}_0$.

The only other input left is the trajectory itself.
The geodesic that begins at rest at $r\rightarrow \infty$ has
\begin{equation}
  \begin{split}
  \underline{U}_0(\tau) &= \lambda^{-1/2}\left(1 - w\right) e^{ w +\frac12 w^2+ \frac13 w^{3}},\\
  \underline{V}_0(\tau) &= \lambda^{1/2}\left(1 + w\right) e^{ -w +\frac12 w^2- \frac13 w^{3}},
\end{split}
\end{equation}
where $w \equiv w(\tau) = \sqrt{\frac{r(\tau)}{2GM}}$ and hence $\frac{dw}{d\tau} = -\frac1{2GM}\frac1{2 w^2}\,$.
Utilizing the coordinate transformation Eq.~\eqref{eq:coord-trans-schwarz-krusk}, we have in Schwarzschild coordinates $U_0(\tau) = - 4GM \log[-\underline{U}_0(\tau)]$ and $V_0(\tau) = 4GM \log[ \underline{V}_0(\tau)]$.
One can then compute the objects
\begin{equation}
  \begin{split}
  \frac{\underline{U}_0'(\tau)}{\underline{U}_0(\tau)} & = \frac{1}{4GM\left(\sqrt{\frac{2GM}r} - 1\right)} = -\frac{U_0'(\tau)}{4GM}\,, \\
  \frac{\underline{V}_0'(\tau)}{\underline{V}_0(\tau)} & =  \frac{1}{4GM\left(\sqrt{\frac{2GM}r} + 1\right) } = \frac{V_0'(\tau)}{4GM}\,.
\end{split}
\end{equation}
This all allows us to write down the full solution
\begin{equation}
  \begin{split}
  F_A^{\mathrm{dyn}} & = \begin{cases}
    -\frac{\left(\sqrt{\frac{2GM}{r}} + 1\right)^2}{48\pi(2GM)^2}\left[ 1+ \frac{4 GM} r + \frac{3(2GM)^2}{r^2} \right], \\ \hspace{55pt} \text{Hawking-Hartle/Unruh vacuum}, \\
    \frac1{12\pi}\frac{1 - \frac34 \frac{2GM}{r}}{\left(\sqrt{\frac{2GM}r} - 1\right)^2}\frac{GM}{r^3}, \\ \hspace{80pt} \text{Boulware vacuum}.
  \end{cases} \\
  F_B^{\mathrm{dyn}} & = \begin{cases}
    -\frac{\left(\sqrt{\frac{2GM}{r}} - 1\right)^2}{48\pi(2GM)^2}\left[ 1+ \frac{4 GM} r + \frac{3(2GM)^2}{r^2} \right], \\ \hspace{80pt} \text{Hawking-Hartle vacuum}, \\
    \frac1{12\pi}\frac{1 - \frac34 \frac{2GM}{r}}{\left(\sqrt{\frac{2GM}r} + 1\right)^2}\frac{GM}{r^3}, \\ \hspace{80pt} \text{Boulware/Unruh vacuum}.
  \end{cases}
\end{split}
\end{equation}
(We will explain the allusions to the Unruh vacuum case shortly.)
A note on the signs: for either plate, a negative force indicates a force directed towards the center of the apparatus while a positive force indicates a force away from the apparatus's center of mass.

Thus the Hartle--Hawking vacuum pushes the apparatus away from the black-hole while also pushing the plates together, and the Boulware vacuum tries to pull the plates towards the black-hole while also pulling the plates apart.

Furthermore, $F_A^{\mathrm{dyn}}$ diverges at the horizon in the Boulware vacuum.
In the case of a star that has not collapsed, the Boulware state should apply on both sides of the apparatus up until the apparatus lands on the star.
When a horizon exists, however, Boulware conditions in that region are physically implausible.

On the other hand, for an eternal black hole we would expect the (more regular) radiation pressure induced by the Hawking--Hartle vacuum to apply on plate $A$ as the apparatus falls towards the horizon. 

Last, we consider the Unruh vacuum given by the metric long after a black hole collapse \cite{unruhNotesBlackholeEvaporation1976,daviesEnergymomentumTensorEvaporating1976}.
In this case, we have new coordinates $(u_c,v_c)$ which respect
\begin{equation}
    ds^2 = \left( 1 - \frac{2M}{r}\right)\left[ \frac{4M}{A-u_c} + \cdots \right] du_c dv_c,
\end{equation}
we note that $\Gamma\indices{^v_{vv}}$ will remain unchanged from the Boulware vacuum and thus $F_B^{\mathrm{dyn}}$ will similarly be left unaltered. 
It is the force on plate $A$ which differs from Boulware, but in fact $u = -4M \log(C - u_c) + D(u_c)$, where $D(u_c)$ is slowly varying and collapse-dependent \cite{daviesEnergymomentumTensorEvaporating1976}.
The dominant part is related to Kruskal coordinates on only the $u$ coordinate such that $u_c \approx \underline{u}$ while $v_c = v$ remains unaltered. 
Therefore, $\Gamma\indices{^{u_c}_{u_c u_c}}$ matches that of the Hawking-Hartle vacuum, and therefore the collapse causes a radiation pressure on the plate nearer the black hole while the plate further away from the hole has the same force as the Boulware vacuum.

This allows for an intuitive explanation of the forces that we see.
In the Hawking-Hartle vacuum, the black hole is in thermal equilibrium with radiation coming from $r=+\infty$; Hawking radiation leaves the black hole while radiation comes in to balance it from $r=+\infty$. 
Indeed $F_A^{\mathrm{dyn}}<0$ indicates radiation pressure away from the black hole while $F_B^{\mathrm{dyn}}<0$ indicates pressure towards the black hole from the radiation from $r=\infty$.
In the collapsing star case, the black hole is not in thermal equilibrium so we instead find $F_B^{\mathrm{dyn}}>0$, the Boulware result.

Last, we note that in the Unruh vacuum the apparatus is unequivocally ``pushed'' away from the black hole by dynamical forces ($F_A^\mathrm{dyn}$ and $F_A^\mathrm{dyn}$ now have their magnitudes added together instead of subtracted); not only does the black hole not ``suck'' the apparatus in, it tries to push it away \footnote{This ``push'' is very weak and would only be a slight deviation from the geodesic motion of the plate.}.

\section{Conclusions}

Using properties of free conformal field theory, we have been able to show how the Casimir \emph{force} and energy changes on plates that are suddenly put into free fall.
Firstly, the curvature of space redistributes the energy between the plates in a tidal manner, in analogy with a fluid in a container.
This leads to an increase in the Casimir force as negative energy gets pushed to the edges of the apparatus.
Secondly, the field  between the plates can be put out of equilibrium and begin to slosh back and forth between the plates, causing changes in the forces experienced by the plates.
Although excitations are being created between the plates, we see an increase in negative energy density near the plates and a corresponding increase in attraction between the plates.

The full calculation also includes radiation pressure outside of the apparatus, which contributes to the force; particulars of the system then indicate whether the radiation pressure pulls the plates together or pushes them apart.  In particular, in the example of a spherical body we see a difference in radiation pressure depending on whether the body is a star or a black hole.
The star pulls the Casimir apparatus closer and stretches it by radiation pressure, while the black hole tends to push the apparatus away and compress it.

From an observational point of view, a real scalar field in curved space appears naturally in a superfluid system \cite{unruhExperimentalBlackHoleEvaporation1981,garaySonicBlackHoles2001a} as a phonon field in the acoustic limit.
If one considers objects that interact with that phonon field, one gets a Casimir force between them \cite{schecterPhononMediatedCasimirInteraction2014}, and if that superfluid is flowing, the phonons can be described with a curved space background \cite{unruhExperimentalBlackHoleEvaporation1981,visserAcousticBlackHoles1998a}.
Therefore, in a flowing superfluid, the effects described here should occur.
While it is beyond the scope of this work to explore experimental possibilities, we note this as a potential avenue for future work.

Further, there has also been considerable work involving entanglement \cite{dodonovQuantumHarmonicOscillator2005} of photon modes caused by the motion of a cavity such as the apparatus described here.
The methods and implications of this work might find harbor within that community as well.

All of the results here are determined in the context of 1+1D free field theory.
In 3+1D, two of us have also considered a falling Casimir apparatus \cite{sorgeCasimirEffectFreefall2019} where the changing (time-dependent) gravitational field leads to corrections to the Casimir energy in addition to dynamical effects.
However, that work inherently could not look at tidal effects and the full dynamical response, something that the techniques in 1+1D allow.
We expect that the tidal and nonequilibrium effects persist into higher dimensions, but that must be left to future work.
However, the results are provocative: The Casimir energy itself is behaving in many respects as a classical fluid in curved space both in and out of equilibrium.

\section*{Acknowledgments}

We thank Victor Galitski for discussions that led to this work. We also thank Gil Refael and Yoni Bentov for helpful discussions. JHW thanks the Air Force Office for Scientific Research for support.
JHW performed part of this work at the Aspen Center for Physics, which is supported by National Science Foundation grant PHY-1607611.

\appendix

\section{Perturbation theory} \label{sec:perturbation-theory}

The setup, as described in the main text, is a set of paths $P_\eta = (U_\eta(\tau),V_\eta(\tau))$, where $(U_0(\tau),V_0(\tau))$ is a timelike geodesic.
The theory described below works for timelike worldlines that are not geodesics, but the math becomes onerous and little insight is gained from that analysis.
Therefore, to restate our conditions on $P_0(\tau)$,
\begin{equation}
  \begin{split}
  C(P_0) U_0'(\tau)V_0'(\tau) &= 1, \\
  U_0''(\tau) + \Gamma\indices{^u_{uu}} U_0'(\tau)^2 & = 0, \\
  V_0''(\tau) + \Gamma\indices{^v_{vv}} V_0'(\tau)^2 & = 0.
\end{split}
\end{equation}
The  curves with $\eta\neq 0$ are defined as being a fixed distance $\eta$ from $P_0$\,.
In other words, they solve
\begin{equation}
  \begin{split}
  \partial_\eta^2 U_\eta(\tau) + \Gamma\indices{^{u}_{uu}} [\partial_\eta U_\eta(\tau)]^2 & = 0, \\
  \partial_\eta^2 V_\eta(\tau) + \Gamma\indices{^{v}_{vv}} [\partial_\eta V_\eta(\tau)]^2 & = 0,
\end{split}
\end{equation}
with the initial conditions
\begin{equation}
  \begin{split}
  (U_\eta(\tau), V_\eta(\tau))|_{\eta=0} &= (U_0(\tau), V_0(\tau)), \\
  (\partial_{\eta} U_\eta(\tau), \partial_\eta V_\eta(\tau))|_{\eta=0} &= (-U_0(\tau), V_0(\tau)). \label{eq:spacelike-geodesic}
\end{split}
\end{equation}
The second condition [Eq.~\eqref{eq:spacelike-geodesic}] implies that $\eta$ parametrizes a spacelike geodesic by proper distance.
Further, in dimension~2 it describes a spacelike surface orthogonal to the proper time of the trajectory $P_0\,$.
For our purposes we will need all second and third derivatives, which can be calculated as
\begin{equation}
  \begin{split}
  \partial_{\tau}\partial_{\eta} U_\eta(\tau)|_{\eta=0} & = \Gamma\indices{^u_{uu}} U'_0(\tau)^2, \\
  \partial_{\tau}\partial_{\eta} V_\eta(\tau)|_{\eta=0} & = -\Gamma\indices{^v_{vv}} V'_0(\tau)^2,
\end{split}
\end{equation}
and
\begin{multline}
\partial_{\eta}^3 U_{\eta}(\tau)|_{\eta=0}  = [\partial_u\Gamma\indices{^u_{uu}} - 2 (\Gamma\indices{^u_{uu}} )^2 ]U_0'(\tau)^3 \\+ \frac14 R U_0'(\tau),
\end{multline}
\begin{multline}
   U_0'''(\tau) = - \partial_{\tau}^2 \partial_{\eta} U_{\eta}(\tau)|_{\eta=0} = \partial_\tau \partial_{\eta}^2 U_{\eta}(\tau)|_{\eta=0} = \\ -[\partial_u\Gamma\indices{^u_{uu}} - 2 (\Gamma\indices{^u_{uu}} )^2 ]U_0'(\tau)^3 + \frac14 R U_0'(\tau).
\end{multline}
\begin{multline}
\partial_{\eta}^3 V_{\eta}(\tau)|_{\eta=0}  = -[\partial_v\Gamma\indices{^v_{vv}} - 2 (\Gamma\indices{^v_{vv}} )^2 ]V_0'(\tau)^3 \\ - \frac14 R V_0'(\tau),
\end{multline}
\begin{multline}
   V_0'''(\tau) = \partial_{\tau}^2 \partial_{\eta} V_{\eta}(\tau)|_{\eta=0} = \partial_\tau \partial_{\eta}^2 V_{\eta}(\tau)|_{\eta=0} = \\ -[\partial_v\Gamma\indices{^v_{vv}} - 2 (\Gamma\indices{^v_{vv}} )^2 ]V_0'(\tau)^3 + \frac14 R V_0'(\tau). \label{eq:geodesic-derivative}
\end{multline}

We place the plates at $A=-L/2$ and $B=L/2$ so that they are separated by a proper distance $L$, and we can then write  Eq.~\eqref{eq:H_solve} suggestively as 
\begin{equation}
    H\circ V_B \circ U_B^{-1}(u) = H\circ V_A \circ U_A^{-1}(u) + 2L
\end{equation}
or in terms of the central proper time
\begin{multline}
  H\circ V_0 \circ V_0^{-1} \circ V_{L/2}\circ U_{L/2}^{-1} \circ U_0(\tau)  \\
  = H \circ V_0 \circ V_0^{-1} \circ V_{-L/2} \circ U_{-L/2}^{-1} \circ U_0(\tau) + 2L. \label{eq:modified_Hsolv}
\end{multline}
With this form, we can calculate the objects that look like
\begin{equation}
  U_{\eta}^{-1}\circ U_0(\tau) = \tau'.
\end{equation}
Assuming that $\tau' = \tau + \alpha \eta + \beta \eta^2 + \gamma \eta^3 + \cdots$, we can expand
\begin{equation}
  U_0(\tau) = U_{\eta}(\tau'),
\end{equation}
so that if we use the notation $\partial_\eta^n U_0(\tau) \equiv \partial_\eta^n U_\eta(\tau)|_{\eta=0}\, $,
\begin{multline}
  0 = [\alpha U'_0(\tau) + \partial_\eta U_0(\tau)]\eta \\+ \left[\beta U'_0(\tau) + \tfrac12 \alpha^2 U''_0(\tau) + \alpha \partial_\eta U_0'(\tau) + \tfrac12 \partial_\eta^2 U_0(\tau)\right] \eta^2 \\ + \left[\gamma U_0'(\tau) + \alpha \beta U_0''(\tau) + \beta \partial_\eta U_0'(\tau) + \tfrac16\alpha^3 U_0'''(\tau) \right. \\  \left. +\tfrac12 \alpha^2 \partial_\eta U_0''(\tau) + \tfrac12 \alpha \partial_\eta^2 U_0'(\tau) + \tfrac16 \partial_\eta^3 U_0(\tau)\right]\eta^3 + \cdots.
\end{multline}
Solving each of these, order-by-order, we get $\alpha=1$, $\beta=0$, and $\gamma = -\frac1{12} R$.
Therefore, we have
\begin{equation}
  \begin{split}
 U_\eta^{-1} \circ U_0(\tau) & = \tau + \eta - \tfrac1{12} R \eta^3 + \cdots, \\
 U_0^{-1} \circ U_\eta(\tau) & = \tau - \eta + \tfrac1{12} R \eta^3 + \cdots.
\end{split}
\end{equation}
Being careful with minus signs, we can similarly find
\begin{equation}
  \begin{split}
  V_\eta^{-1} \circ V_0(\tau) & = \tau - \eta + \tfrac1{12} R \eta^3 + \cdots, \\
  V_0^{-1}\circ V_\eta(\tau) & = \tau + \eta - \tfrac1{12} R \eta^3 + \cdots.
\end{split}
\end{equation}
Therefore, we can simplify Eq.~\eqref{eq:modified_Hsolv} into
\begin{multline}
 H \circ V_0\bigl(\tau + L - \tfrac1{48} R L^3\bigr)  \\= H\circ V_0\bigl(\tau - L + \tfrac1{48} R L^3\bigr) + 2L.
\end{multline}
The curvature $R = R[P_0(\tau)]$ is $\tau$-dependent, so for $\tau = \tau_0 + \Delta \tau$, we can expand $R = R(\tau_0) + \partial_\tau R(\tau_0) \Delta \tau $, and we have, for instance, $H\circ V_0[ \Delta\tau(1 - \frac1{48}\partial_\tau R(\tau_0)L^3) + \tau_0 + L - \tfrac1{48} R(\tau_0) L^3]$.
We can then find a perturbative solution in $\Delta \tau$ which is simply
\begin{multline}
  H \circ V_0(\tau_0+\Delta \tau) \\ = H\circ V_0(\tau_0) + \Delta\tau \left(1 + \frac1{48} R L^2 \right) + O(\Delta \tau^2).
\end{multline}
This expansion allows us to directly read off the derivative of $H$ as
\begin{equation}
  H'(v) = \frac1{V_0'\circ V_0^{-1}(v)} \left(1 + \frac1{48} R L^2 + O(L^4) \right),
\end{equation}
where the order $L^4$ comes from a careful analysis of Eq.~\eqref{eq:modified_Hsolv}.
Additionally, one can obtain in a similar manner
\begin{equation}
  (H\circ p)'(u) =  \frac1{U_0'\circ U_0^{-1}(u)} \left(1 + \frac1{48} R L^2 + O(L^4) \right).
\end{equation}

\section{Initial conditions}\label{sec:initial_conditions}

In order to find the appropriate $Q$ transformation from Fig.~\ref{fig:coordinate-transformations} and Fig.~\ref{fig:InitialCond_plates}, we make some assumptions about the initial state.
In order to apply  Eq.~\eqref{eq:Htilde_soln}, we need to assume that plate $B$ is at both  coordinate and proper distance $L$ from plate $A$.
Taken precisely: for $t<0$ we assume $U_A(\tau) - V_A(\tau) = 0$ and $V_B(\tau)-U_B(\tau) = 2L$ where $L$ is the proper distance (but $x\equiv \frac12(v - u)$ is \emph{not} necessarily the proper distance; that is to say that when $x=L$ it coincides with the proper distance but when $x\neq L$ it may not correspond).
As we describe in the main text, this assumption is done without loss of generality, but applications of this theory must be scaled appropriately.
Next, we assume that the metric initially has a timelike Killing vector $\partial_t$ so that $C(u,v) = C(v-u)$ and the center of the apparatus begins ``falling'' at $t=0$ and $\tau = 0$.

Lastly, for ease we define $r(v) \equiv \{H(v)\}_{t_1 - L}^{t_1 + L}$ and so there exists an $n$ such that
\begin{equation}
    \tilde H(v) = 2 n L + \begin{cases}
  (H\circ p)^{-1}[r(v)], & r(v) < 0, \\
  H^{-1}[r(v)], & r(v) > 0.
  \end{cases}
\end{equation}

Under these conditions, both plates also begin moving at $t=0$, so we have $t_0=0$ in particular.
This can be understood since the trajectory of the center of the plates before being dropped has the 2-velocity $C^{-1/2}(2x_0)(1,1)$ which when parallel transported along $x = \frac12(v-u)$ is just $C^{-1/2}(2x)(1,1)$.
Therefore, $V_\eta(0) = - U_\eta(0)$ both before and after $t=0$; this immediately implies that both $t_0=0=t_1$.

We now need to determine where everything is in space with respect to the proper distance $L$ between the plates.
We define the observer's coordinate position as $x_0$, and the geometric quantities at that position are
\begin{equation}
  \begin{split}
  C_0(x_0) & = C(2x_0) \\
   \Gamma_0 & = \frac12\partial_{x_0} \log C_0(x_0) ,\\
   R_0 & = \frac{\partial_{x_0}^2 \log C_0(x_0)}{C_0(x_0)}\,.
 \end{split}
\end{equation}
To determine distances, we will need the expansion
\begin{multline}
  \sqrt{C(2x )}
  =  \sqrt{C_0}\left( 1 + \Gamma_0(x - x_0) + \right. \\ \left. \tfrac14(C_0 R_0 + 2\Gamma_0^2)(x-x_0)^2 \right) + \cdots.
\end{multline}
One can easily determine now where $x_0$ is by considering
\begin{equation}
  \begin{split}
   \frac L2 & = \int_0^{x_0} \sqrt{C(2x)} dx, \label{eq:midwayx0} \\
    \frac{L}{2\sqrt{C_0}} & = x_0 - \frac12\Gamma_0 x_0^2 + \frac1{12}(R_0 C_0 + 2 \Gamma_0^2) x_0^3  + \cdots.
  \end{split}
\end{equation}
This series can be inverted to give
\begin{multline}
  x_0 = \frac{L}{2\sqrt{C_0}} + \frac12\Gamma_0 \left( \frac{L}{2\sqrt{C_0}} \right)^2 \\ +\frac1{12}( 4 \Gamma_0^2- R_0 C_0)\left( \frac{L}{2\sqrt{C_0}}\right)^3 + \cdots.
\end{multline}
To enforce the constraint on the position of plate $B$, we have
\begin{equation}
  \frac{L}2 = \int_{x_0}^L \sqrt{C(2x)}\, dx,
\end{equation}
and we obtain
\begin{equation}
  C_0 = 1 + \tfrac1{24} \left(4\Gamma_0^2 - R_0 \right) L^2 + O(L^4).
\end{equation}
This lets us simplify to
\begin{equation}
  x_0 = \frac L2 + \frac12 \Gamma_0 \left(\frac L 2 \right)^2 + O(L^4).\label{eq:x0equation}
\end{equation}
With this setup, we can now determine $Q$ for when we drop these plates.

Dropping the plates  amounts to setting the initial conditions on $U_0(0) = -x_0$, $V_0(0) = x_0$ and
\begin{equation*}
  \begin{split}
U'_0(0) & = V'_0(0) = \frac1{\sqrt{C_0}} \\ & = 1 - \frac1{48} \left(4\Gamma_0^2 - R_0\right) L^2 + O(L^4).
\end{split}
\end{equation*}
The geodesic equations are particularly simple at this point  as well:
\begin{equation}
  \begin{split}
  U_0''(0) & = \Gamma_0 U_0'(0)^2, \\
  V_0''(0) & = -\Gamma_0 V_0'(0)^2, \\
  U_0'''(0) & = 2\Gamma_0^2 U_0'(0)^3, \\
  V_0'''(0) & = 2\Gamma_0^2 V_0'(0)^3.
\end{split}
\end{equation}

With all of this established, we can now take the inverses of $H(v)$ and $H\circ p(u)$.
We know that $H(0)=0$ and $H\circ p(0)=0$, so we can expand the functions about that point.
Partially resumming $H$ after Taylor-expanding and using Eq.~\eqref{eq:Hpert}  gives
\begin{equation}
  H(v) = [V_0^{-1}(v)- V_0^{-1}(0)]\left(1 + \frac1{48} R_0 L^2 \right),
\end{equation}
for small $v$.
Therefore, we have two equations
\begin{align}
  H^{-1}[r( v)]& =  V_0\left[ V_0^{-1}(0) + r( v)\left(1 - \tfrac1{48} R_0  L^2\right)\right], \label{eq:Hinv}\\
(H\circ p)^{-1}[r( v)] & =  U_0\left[ U_0^{-1}(0) + r( v)\left(1 - \tfrac1{48} R_0  L^2\right)\right].\label{eq:Hcircpinv}
\end{align}
We  expand $ V_0^{-1}(0)$ by considering $ V_0^{-1}(0) =  V_0^{-1}[V(0) - x_0]$, and similarly for $ U_0^{-1}(0)$:
\begin{equation}
  \begin{split}
  V_0^{-1}(0) & = -\frac L2 + \frac{R_0}{12} \left( \frac L 2\right)^3 + \cdots, \\
  U_0^{-1}(0) & = \frac L2 - \frac{R_0}{12} \left( \frac L 2\right)^3 + \cdots.
\end{split}
\end{equation}

With all of this we can expand the inverse functions that we require:
\begin{multline}
  H^{-1}[r(v)]  = \left(1 + \frac12\Gamma_0 \, L + \frac1{6} [\Gamma_0 L]^2  \right) r(v) \\ - \frac12\left(1 +\Gamma_0 L  \right)\Gamma_0 r(v)^2 \\
  + \frac1{3} \Gamma_0^2 \, r(v)^3 + \cdots,
\end{multline}
and similarly
\begin{multline}
  (H\circ p)^{-1}[r(v)]  = \left(1 + \frac12\Gamma_0 \, L + \frac1{6} [\Gamma_0 L]^2  \right) r(v) \\ + \frac12\left(1 + \Gamma_0 L  \right)\Gamma_0 r(v)^2 \\
  + \frac1{3} \Gamma_0^2 r(v)^3 + \cdots.
\end{multline}

As we show in the main text, these expansions define a periodic function that is continuous, but its second derivative is not.
For completeness, we write down the whole function:
\begin{multline}
  \tilde H(v) = H(v) + \left(\frac12\Gamma_0 L + \frac1{6} [\Gamma_0 L]^2  \right) r(v) \\ - \mathrm{sgn}[r(v)] \frac12\left(1 + \Gamma_0 L  \right)\Gamma_0 \, r(v)^2 \\
  + \frac1{3} \Gamma_0^2 \, r(v)^3 + \cdots.
\end{multline}
Finally, note that this is true in a starting coordinate system that has been scaled so that plate $A$ is at $x=0$  and plate $B$ is at $x=L$.
The object $\Gamma_0$ is dependent on this scaling, so we need to be careful when applying this formula.

\bibliography{references.bib}

\end{document}